\documentclass[english, twocolumn]{aa}
\usepackage[varg]{txfonts}
\usepackage{graphicx}
\usepackage{mathrsfs}
\usepackage{amsmath}
\usepackage{txfonts}
\usepackage{enumerate}
\usepackage{caption}
\usepackage{multirow}
\usepackage{subcaption}
\usepackage[colorlinks=true, citecolor=blue]{hyperref}
\usepackage{bm} 
\usepackage{babel}
\usepackage[switch]{lineno}
 \usepackage{natbib}
\bibpunct{(}{)}{;}{a}{}{,} % to follow the A&A style
\graphicspath{{./}}
\usepackage{etoolbox}
\usepackage{xcolor}  %for comment color
\usepackage{orcidlink}

\begin{document} 
% \linenumbers

\author{
L.~Giunti\,\orcidlink{0000-0002-3395-3647}\,\inst{\ref{APC}} 
\and F. Acero\,\orcidlink{0000-0002-6606-2816}\,\inst{\ref{IRFU}}
\and B.~Kh\'elifi\,\orcidlink{0000-0001-6876-5577}\,\inst{\ref{APC}}
\and K.~Kosack\,\orcidlink{0000-0001-8424-3621}\,\inst{\ref{IRFU}}
\and A.~Lemi\`ere\,\orcidlink{0000-0002-6682-7188}\,\inst{\ref{APC}}
\and R.~Terrier\,\orcidlink{0000-0002-8219-4667}\,\inst{\ref{APC}}
}

\defcitealias{Giunti}{A21}
\defcitealias{fujinaga}{F11}

\institute{
Astroparticule et Cosmologie, Université de Paris Cit\'e, CNRS/IN2P3, F-75013 Paris, France \label{APC}
\and Université Paris-Saclay, Université Paris Cité, CEA, CNRS, AIM de Paris-Saclay, 91191 Gif sur Yvette
 \label{IRFU}
}

%% Redefine numeric symbols for footnotes
\makeatletter
\renewcommand*{\@fnsymbol}[1]{\ifcase#1\or*\or$\dagger$\or$\ddagger$\or**\or$\dagger\dagger$\or$\ddagger\ddagger$\fi}
\makeatother

%%%%%%%%%%%%%%%%%%%%%%%%%%%%%%%%%%%%%%%%
\date{Received ?; Accepted ?}

\newcount\Comments  % 0 suppresses notes to selves in text
\Comments=1   % TODO: set to 0 for final version
\newcommand{\kibitz}[2]{\ifnum\Comments=1\textcolor{#1}{#2}\fi}
\definecolor{darkgreen}{rgb}{0,0.5,0}
\newcommand{\fabio}[1]{\kibitz{darkgreen}{[FA: #1]}}

\newcommand{\SEC}{\,\mathrm{s}\,}
\newcommand{\SECminusone}{\,\mathrm{s}^{-1}\,}
\newcommand{\EV}{\,\mathrm{eV}\,}
\newcommand{\MEV}{\,\mathrm{MeV}\,}
\newcommand{\GEV}{\,\mathrm{GeV}\,}
\newcommand{\TEV}{\,\mathrm{TeV}\,}
\newcommand{\PEV}{\,\mathrm{PeV}\,}
\newcommand{\TEVtwo}{\,\mathrm{TeV}^2\,}
\newcommand{\TEVminustwo}{\,\mathrm{TeV}^{-2}\,}
\newcommand{\ERG}{\,\mathrm{erg}\,}
\newcommand{\ERGtwo}{\,\mathrm{erg}^2\,}
\newcommand{\ERGminustwo}{\,\mathrm{erg}^{-2}\,}
\newcommand{\CMminustwo}{\,\mathrm{cm}^{-2}\,}
\newcommand{\CMminusthree}{\,\mathrm{cm}^{-3}\,}
\newcommand{\FLUXUNIT}{\,\mathrm{erg}\, \mathrm{cm}^{-2}\, \mathrm{s}^{-1}\,}
\newcommand{\jsev}{\mbox{HESS J1702-420} }
\newcommand{\jsevA}{\mbox{HESS J1702-420A} }
\newcommand{\jsevB}{\mbox{HESS J1702-420B} }
\newcommand{\jsevvirg}{\mbox{HESS J1702-420}, }
\newcommand{\jsevpv}{\mbox{HESS J1702-420}; }
\newcommand{\jsevdot}{\mbox{HESS J1702-420}. }
\newcommand{\jsevAvirg}{\mbox{HESS J1702-420A}, }
\newcommand{\jsevBvirg}{\mbox{HESS J1702-420B}, }
\newcommand{\jsevAdot}{\mbox{HESS J1702-420A}. }
\newcommand{\jsevBdot}{\mbox{HESS J1702-420B}. }
\newcommand{\ddeg}{^{\,\text{o}}}
\newcommand{\gr}{\mbox{$\gamma$-ray} }
\newcommand{\grvirg}{\mbox{$\gamma$-ray}, }
\newcommand{\Suzaku}{\emph{Suzaku} }
\newcommand{\XMM}{\emph{XMM-Newton} }

\newcommand{\gp}{\emph{Gammapy} }
\newcommand{\banana}{\mbox{XMMU J170147.3-421407} }
\newcommand{\bananav}{\mbox{XMMU J170147.3-421407}, }
\newcommand{\bananap}{\mbox{XMMU J170147.3-421407}. }

  \title{Constraining leptonic emission scenarios for the PeVatron candidate \jsev with deep \XMM observations}
  \titlerunning{\XMM X-ray counterparts of \jsev}
  \abstract{}
  {\small  We aim to search for a hidden leptonic accelerator, such as a high-$\dot{\mathrm{E}}$ pulsar,   associated with the unidentified TeV object \jsevAdot}
  {\small A 72$\,$ks X-ray observation was carried out with the \XMM satellite, and the resulting data were analyzed jointly with the  publicly available H.E.S.S.\ spectral energy distribution (SED) to  derive constraints on the leptonic contribution to the TeV emission of \jsevAdot   A set of  scripts  dedicated to the multi-wavelength modeling of X-ray and \gr data, based on   \emph{Gammapy}, \emph{Naima} and \emph{Xspec},  was developed in the context of this work and  is made publicly available along with this paper.}
  {\small No object clearly associated with \jsevA was found in the \XMM data. After excluding   the unidentified object \Suzaku src B as a possible X-ray counterpart, and classifying it as a new cataclysmic variable source candidate, strict upper limits on the level of diffuse X-ray emission in the \mbox{HESS J1702-420A} region were derived: $F(2-10\,\text{keV})\lesssim5.4\,\times10^{-5}\,\text{keV}\,\CMminustwo\SECminusone$ at $2\sigma$ ($\approx95.5\%$) confidence level. A tight constraint on the magnetic field was derived, under a one-zone leptonic scenario, by jointly fitting
the \XMM spectra and the H.E.S.S.\ SED:  $B\lesssim 1.45\,\mu$G at  $2\sigma$ level. We additionally report the serendipitous discovery of a new  extended X-ray source with a hard spectral index of $1.99\pm0.45$, named  \bananav that is likely Galactic. Its classification as a high-speed runaway pulsar wind nebula (PWN), possibly associated with \jsevAvirg is not obvious but cannot be  ruled out either.}
  {The hard \gr object \jsevA remains unidentified, but the absence of a clear X-ray counterpart strongly challenges simple leptonic scenarios. The only remaining possible  leptonic counterpart for \jsevA appears to be a newly discovered X-ray source  with extended morphology and hard spectral index, that could be a PWN powered by a high-speed runaway pulsar. }
  %{\small The absence of a clear X-ray counterpart   and the tight magnetic field upper limits that were derived, under one-zone leptonic emission scenarios, significantly strengthen the classification of \jsevA as a hadronic PeVatron candidate. A possible association with a newly discovered extended X-ray source cannot be ruled out with the available data, and should be investigated with further deep multi-wavelength observations.}


\maketitle

%
%-------------------------------------------------------------------
\begin{figure*}
    \centering
    \includegraphics[width=\linewidth]{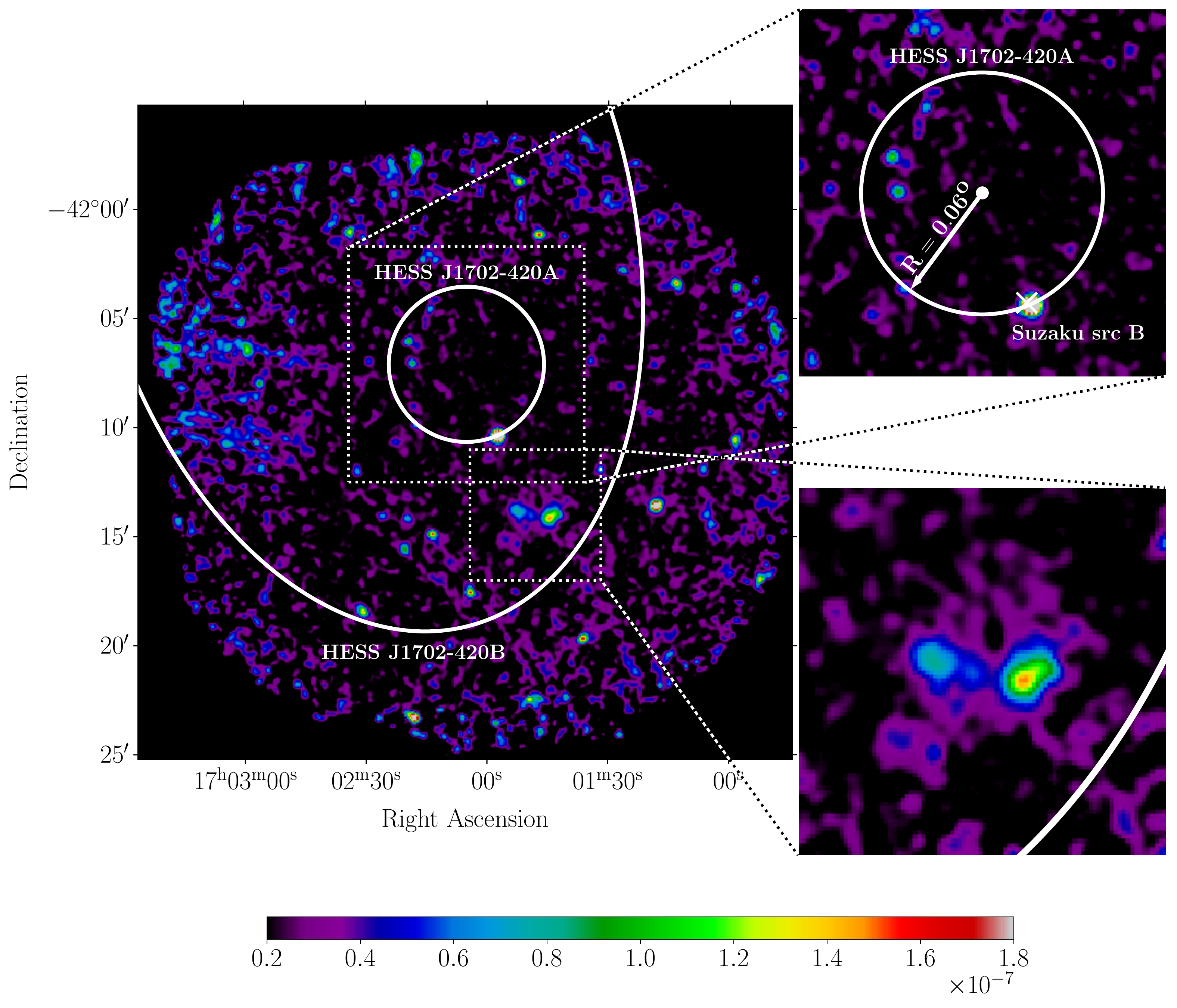}
    \caption{Flux image in the $2-10\,$keV  energy band obtained from the dedicated \XMM   follow-up observation, in the \emph{international celestial reference system} (ICRS) frame. The image is in units of 10$^{-7}$ ph cm$^{-2}$ s$^{-1}$ per $4''\times4''$ pixel, and it has been smoothed with a $\sigma=7''$ Gaussian filter. The TeV morphological models of \jsevA and \jsevB are overlaid.}
    \label{fig1}
\end{figure*}

\section{Introduction}\label{introduction}
\jsev is an extreme particle accelerator discovered during the first H.E.S.S.\ Galactic plane survey campaign~\citep{first_hgps}. Due to its hard power law spectrum and lack of high-energy cutoff, it was soon recognized as a promising candidate accelerator of PeV-energy hadronic cosmic rays, or \emph{PeVatron}. Still, the lack of associated objects at other wavelengths has   hindered   understanding the origin of its TeV \gr emission~\citep{dark, lau}, in particular whether it is produced
by  the inverse-Compton up-scattering of  \mbox{low-energy} photon fields by ultra-relativistic $e^\pm$ (\emph{leptonic} processes) or by  the inelastic collision of  CR hadrons with  interstellar medium nuclei (\emph{hadronic} processes). %This has so far prevented a final judgement on its hadronic PeVatron nature~\citep{dark, Giunti}. 

The H.E.S.S.\ Collaboration has recently published an updated analysis of the region based on deep observations of \jsevvirg which resulted in the discovery of a new small-scale \mbox{($0.06\pm0.02_\text{stat}\pm0.03_\text{syst}\ddeg$ in radius)} emission zone called \jsevA with extremely hard power law spectrum (index of \mbox{$1.53\pm0.19_\text{stat}\pm0.20_\text{syst}$}) above TeV, possibly extending up to 100$\,$TeV (\citet{Giunti}, hereby~\citetalias{Giunti}). If modeled with a simple one-zone hadronic (leptonic) model, this object can be equally well interpreted as an accelerator of protons (electrons) up to at least $0.8\,\PEV$ ($100\,\text{TeV}$)~\citepalias{Giunti}. 

 \citetalias{Giunti}  has pointed out that \jsevA  overlaps with the position of an unidentified X-ray point source called \mbox{\Suzaku src B}, discovered by  \citet{fujinaga} (hereby referred to as \citetalias{fujinaga}). This object has not been properly characterized so far, due to its intrinsic faintness and its location at the border of the \Suzaku field of view in the  \citetalias{fujinaga} pointing.  The discovery of \jsevA has renewed the interest around the nature of \mbox{\Suzaku src B}, since it has opened the possibility of finding the first multi-wavelength association for (at least part of) the TeV emission of \jsevdot 
%In the X-ray domain, deep \emph{Suzaku} observations  the  absence of diffuse X-ray emission in the source region, up to the \Suzaku sensitivity level, and revealed the presence of two extremely faint point-like objects (src A and src B, indicated in Figure~\ref{suzaku}, \emph{right} panel) and the   in the \emph{Suzaku} FoV~\citep{fujinaga}. Due to the poor available statistics and the intrinsic faintness of src A and B, it was not possible to measure their X-ray spectra. This prevented their characterization and and left the question regarding their possible association with \jsev unanswered. 
 We   therefore proposed and obtained a deep observation (72$\,$ks) with the highly sensitive \XMM telescope,  in order to characterize  \mbox{\Suzaku src B}, understand whether it can be a pulsar associated with \jsevA and constrain leptonic TeV emission scenarios based on the presence of a putative low-brightness X-ray  pulsar wind nebula (PWN). 
 
  The \XMM data were processed with a collection of tools including standard X-ray analysis packages, such as  \emph{Xspec} and \emph{Sherpa}~\citep{xspec, sherpa1, sherpa2},  and a new set of Python scripts dedicated to the multi-wavelength fitting of X-ray and \gr data with physically-motivated   models. Our scripts rely on  the  \emph{Gammapy}\footnote{\href{https://gammapy.org/}{https://gammapy.org/}} package~\citep{donath} version 0.19  ~\citep{gammapy0.19} for the X-ray and \gr data handling and fitting in a unified framework. The multi-wavelength spectral modeling can be performed using an arbitrary combination of the functions included in the \emph{Xspec} and \emph{Naima}~\citep{naima} libraries, a task that was not easy to achieve with previously existing tools.  The new scripts, described in Appendix~\ref{gammapyX}, were developed on an open-source basis and are made publicly available along with this paper~\citep{giunti_luca_2022_7092736}, to ensure a full analysis  reproducibility and encourage their application to other studies.
  
  The paper is structured as follows: we start by describing the new X-ray data and their analysis (section~\ref{xmm}),  then we provide a physical interpretation of the  results (section~\ref{discussion}) and conclude with a summary in section~\ref{conclusions}.

 \section{X-ray observations and data analysis}\label{xmm}
 
 To search for possible X-ray counterparts of \jsevAvirg we proposed  a 72 ks
  \XMM observation that was carried out on September 26th 2021. 
  The observation was centered on the position of \mbox{\Suzaku src B}, \mbox{$l=344.093\ddeg$} and \mbox{$b=-0.167\ddeg$} in the Galactic coordinate system~\citepalias{fujinaga}. The region was observed  using the three EPIC instruments on board of the \XMM satellite in Full Frame mode. After flare screening, the remaining exposure time is
63 ks and 44 ks for the MOS and PN cameras, respectively.

 Figure~\ref{fig1} shows the background-subtracted and vignetting-corrected image obtained stacking the MOS1, MOS2 and PN data in the energy band $2-10\,$keV, with two inset plots zooming  on \jsevA (upper right panel) and a newly discovered diffuse  X-ray emission region (bottom right panel). 
 The large scale arc features that pollutes the bottom-left corner of the image are due to straylight (single reflections from a bright X-ray source outside the field of view) but did not affect the analysis of the relevant objects in the central and upper parts of the field of view. 
     % There is no clear counterpart for \jsevA in the XMM data. Therefore we derived an upper limit on the average diffuse X-ray emission and magnetic field corresponding to its TeV size (see section~\ref{});
   %   We report the serendipitous discovery of a new extended X-ray source around the Galactic position $l=344.093 \,\ddeg$ and $b=-0.167 \,\ddeg$. The new object, called here ???, is clearly visible in the ??? panel of figure~\ref{}. Despite the fact that the nature of this object is unclear, an association   with \jsevA is deemed unlikely (see section~\ref{}).
   
 We   focused separately on  \mbox{\Suzaku src B} (section~\ref{suzakuB}), the new extended X-ray source  (section~\ref{banana}) and the \jsevA region (section~\ref{ul}). The regions that were used for spectral extraction and background estimation in all analyses are reported in figure~\ref{fig:xmmregions}. To verify that the analysis systematic uncertainties were under control we tested different region assumptions, obtaining consistent results after taking into account the statistical errors. Because of   high absorption, the spectral analyses were restricted to the $1-10\,$keV energy range. In all cases we used a \emph{Wstat}\footnote{\url{https://heasarc.gsfc.nasa.gov/xanadu/xspec/manual/XSappendixStatistics.html}} fit statistic\footnote{Note that in the case where a background spectra is read, a Wstat is used in \emph{Xspec} but the fit statistic is still labeled Cstat.}, which provides the best likelihood estimates for Poisson-distributed data  with an independent data-driven background estimation.

\subsection{Spectral analysis of Suzaku src B}\label{suzakuB}

\begin{figure*}
    \centering

    \includegraphics[width=0.46\linewidth]{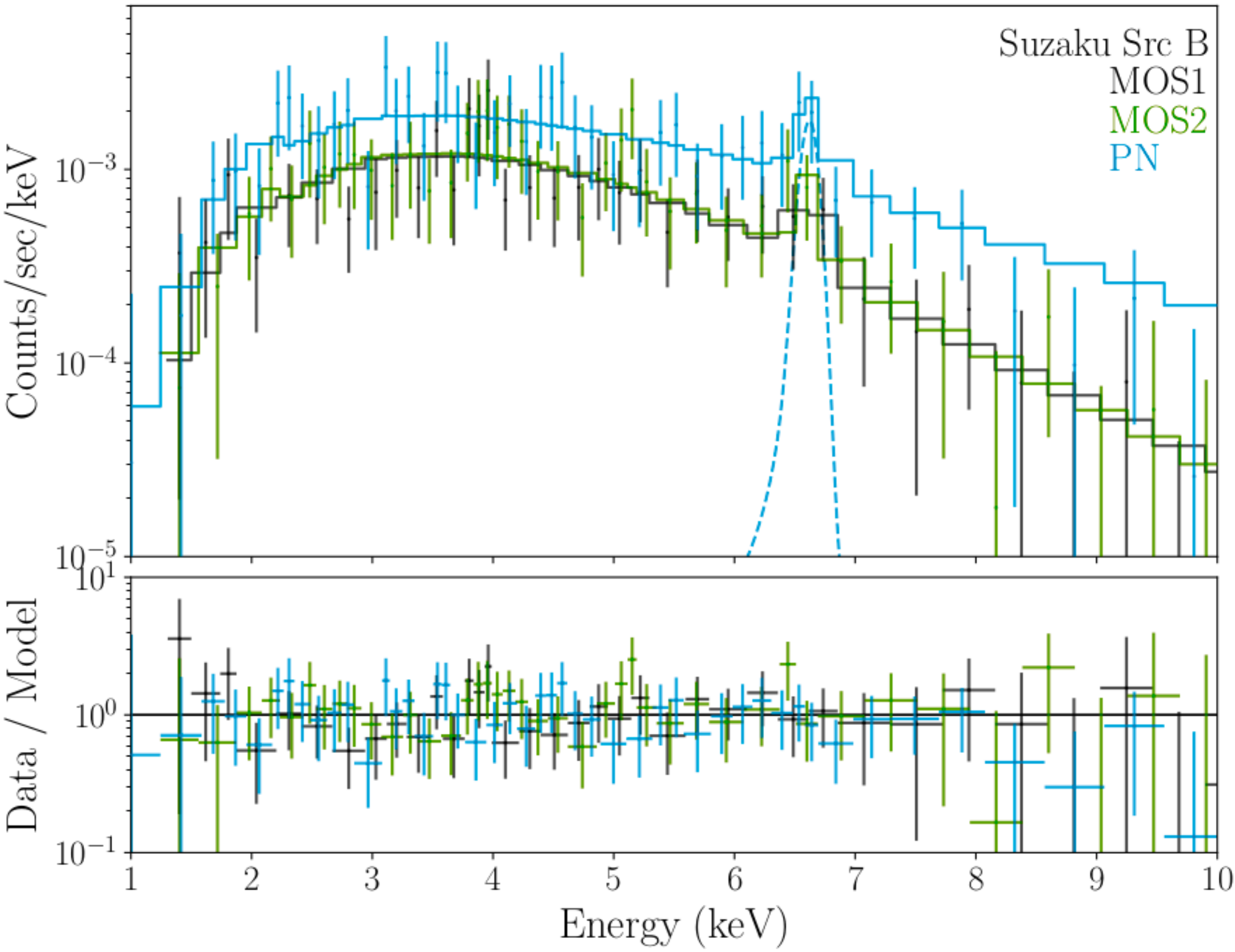} \hspace{0.3cm}    
    \includegraphics[width=0.46\linewidth]{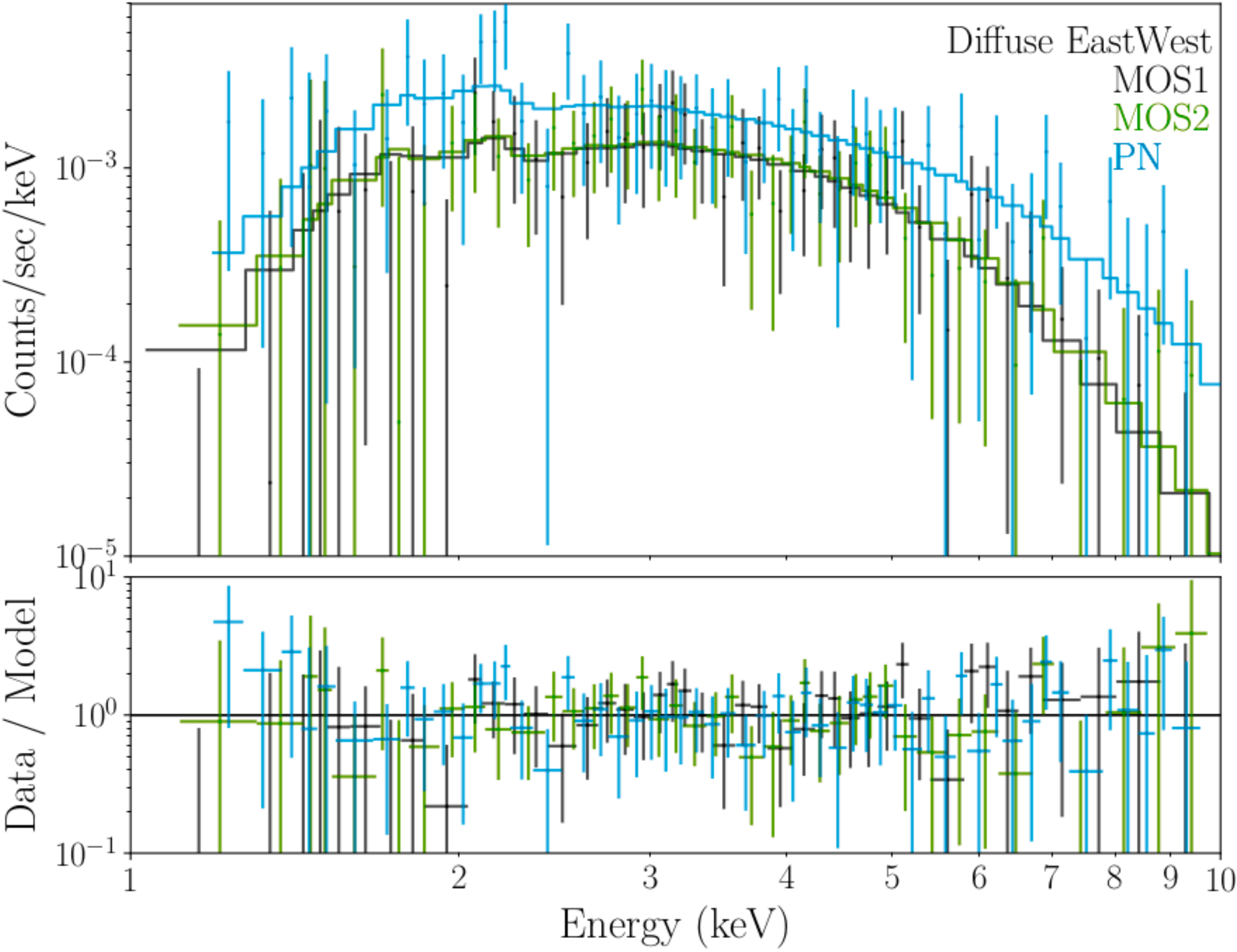}\\
    \includegraphics[width=0.475\linewidth]{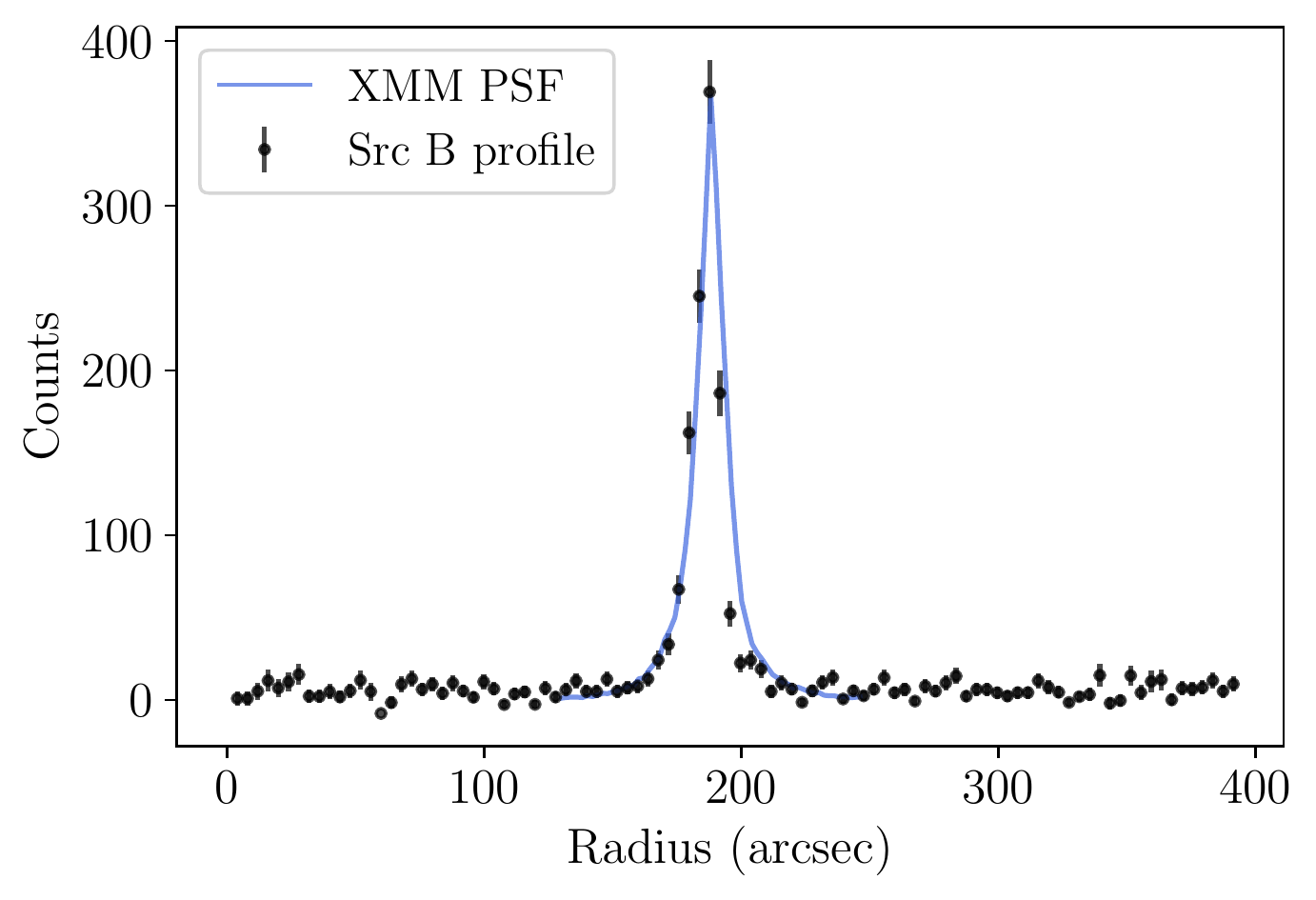} \hspace{0.3cm}
    \includegraphics[width=0.475\linewidth]{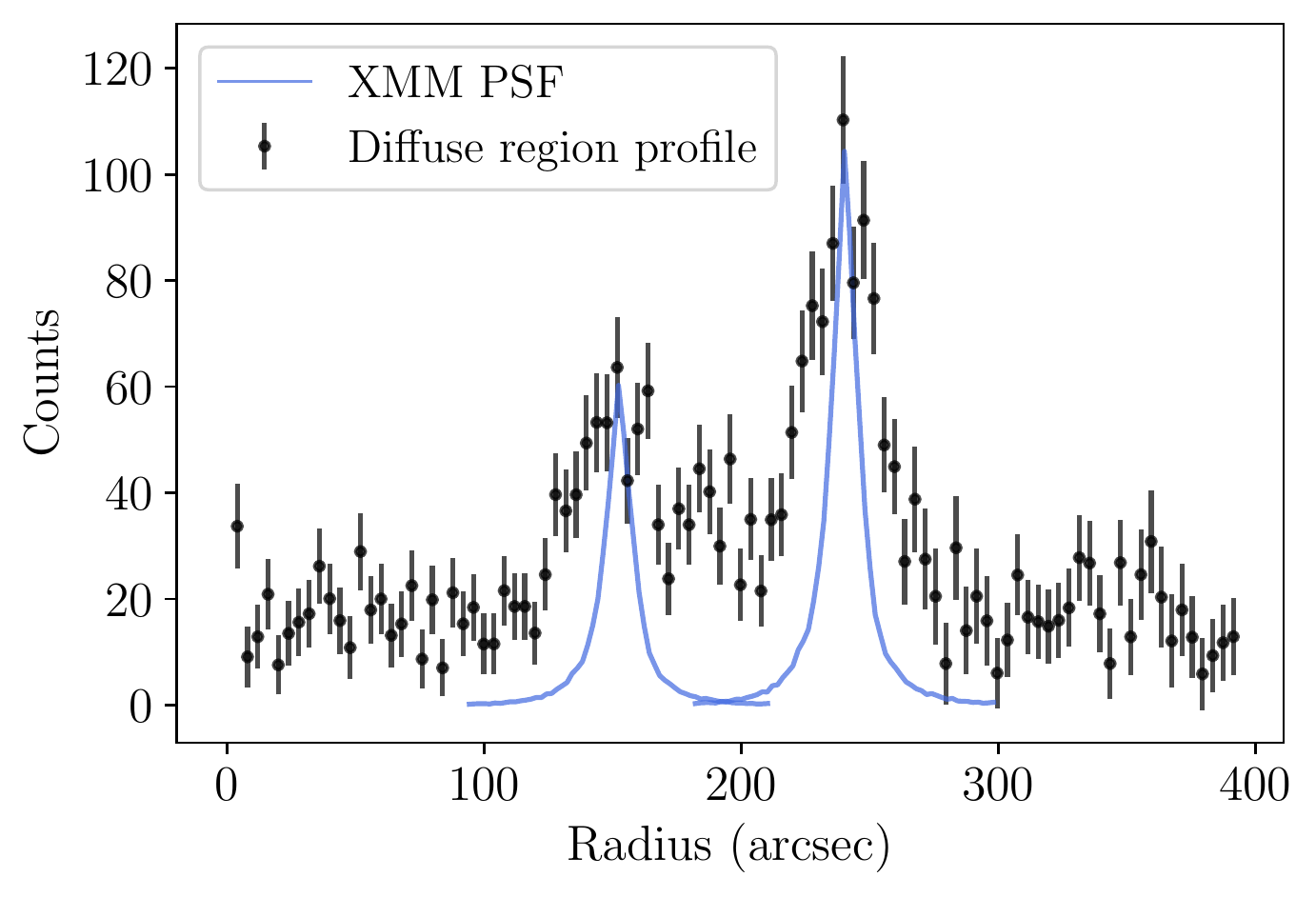}    

    \caption{Spectral and morphological properties of the bright source Suzaku src B and \bananap \textit{Top panel}: \XMM spectra of both sources. The extraction and background regions are shown in figure~\ref{fig:xmmregions}.  All spectra are fitted with an absorbed power-law model except for \Suzaku src B where a Gaussian line is added to model the Fe-K line complex. \emph{On}ly the  Gaussian line for the PN data is shown in the figure. \textit{Bottom panel}: X-ray profiles in the 2-10 keV energy band for both sources, extracted from a $390''$ square box around their centroid position. The PSF was estimated from the same box profile using the 2D calibration PSF images at a reference energy of 3 keV. }
    \label{fig:XMM_spec}
\end{figure*}

\Suzaku src B is a faint and unidentified X-ray point-like source, discovered by \citetalias{fujinaga} with a deep 216$\,$ks \Suzaku pointing and subsequently detected by Chandra (\mbox{2CXO J170157.2-421026}, based on a $\lesssim10\,$ks observation made in 2013). Previous measurements did not provide enough statistics to perform a spectral analysis for this object, making a source classification difficult.  Thanks to the superior effective area of the \XMM satellite, we could for the  first time measure its  spectral properties. 
We used a 15$''$ circular  \emph{On} region (which optimizes the signal to background ratio) to extract the source spectrum, that was modeled  with \emph{Xspec} (version 12.12.0) assuming a \texttt{TbAbs}-absorbed~\citep{Wilms_2000} power law model.

%\begin{equation}\label{abspl}
%    \frac{dN}{dE}=A\left(\frac{E}{E^*}\right)^{-\Gamma} \times\text{exp}\left[-\text{n}_\text{h}\sigma(E)\right]\,,
%\end{equation}
%where $E^*=$<FA> indicates the reference energy, n$_\text{h}$ is the X-ray absorbing column density and $\sigma(E)$ is the photoelectric scattering cross section. 

As shown in figure~\ref{fig:XMM_spec} (top left panel), the X-ray spectrum of \mbox{\Suzaku src B} is well fitted by a heavily absorbed  power law, n$_\text{h}=(5.1\pm 1.1)\times 10^{22}\,\text{cm}^{-2}$,  with a hard spectral index of $1.51 \pm 0.30$.
The presence of an Fe line is visible in the spectrum, especially in the PN camera because of its better effective area at high energies.
When adding a Gaussian to the model, the fit statistics are improved by   $-2\Delta\log (\mathcal{L})=13.5$ (corresponding to a significance of $3.3\sigma$ for 2 degrees of freedom). The best fit line centroid is 6.62 $\pm$ 0.06 keV, while the de-absorbed flux of \mbox{\Suzaku src B}, integrated in the $2-10\,$keV energy band and corrected for the finite emission fraction contained within the \emph{On} region\footnote{For a 15$''$ radius extraction region, the encircled energy fraction is about 70\%. This loss of flux is corrected in the effective area computation using the  \textit{arfgen} tool with the \textit{modelee} parameter.}, is of $2.2 \pm 0.2 \times 10^{-13}\,\text{erg}\,\text{cm}^{-2}\,\text{s}^{-1}$. The bottom left panel of figure~\ref{fig:XMM_spec}
compares the emission profile of \Suzaku src B to the angular resolution of \emph{XMM-Newton}, showing that the source is point-like and   not surrounded by a diffuse X-ray nebula.

%In light of this   analysis, the likelihood of a pulsar-PWN  association between \mbox{\Suzaku src B} and \jsevA is tightly constrained due to several arguments, that are presented in section~\ref{discussion}.

\begin{table}\hspace{-0.5cm}
\begin{tabular}{ l|cccc }

 Parameters & n$_\text{h}$ & Index & Flux (2-10 keV) \\
\hline
\hline
 \Suzaku Src B & 5.1 $\pm$ 1.0 & 1.51 $\pm$ 0.30 & 2.20 $\pm$ 0.2  \\
 Diffuse East  & 2.7 $\pm$ 1.1 & 1.79 $\pm$ 0.63 & 0.48 $\pm$ 0.11   \\
 Diffuse West  & 3.4 $\pm$ 1.1 & 1.63 $\pm$ 0.50 & 0.85 $\pm$ 0.12  \\
 Diffuse East-West  & 3.5 $\pm$ 1.0 & 1.99 $\pm$ 0.45 & 1.25 $\pm$ 0.15  \\
 %\hline
\end{tabular}
\caption{X-ray best-fit spectral parameters for the spectra presented in figure \ref{fig:XMM_spec}. The flux (de-absorbed) is given in units of \mbox{$10^{-13}$ erg cm$^{-2}$ s$^{-1}$} and n$_\text{h}$ in units of $10^{22}\,\text{cm}^{-2}$. Errors are given at the 90\% confidence level.
For \Suzaku Src B, a Gaussian line was added to the model (see main text). }
 \label{tab:specresults}

\end{table}

\subsection{A new diffuse source: \banana}\label{banana}
The new X-ray observation led to the discovery of a  diffuse source located $\approx0.12\ddeg$ away from the center of \jsevAvirg approximately along its $2\sigma$ emission contour (see the bottom right inset of figure~\ref{fig1}). Based on its position, we named this object \bananap  As visible in figure~\ref{fig1}, \banana is strongly asymmetric and composed of two substructures, called here the Eastern and Western lobes, with different shapes and brightnesses. Both lobes are significantly extended in comparison with the \XMM PSF, with apparent sizes of about $1.5'$ (see figure \ref{fig:XMM_spec}, bottom right panel). Due to the low probability of discovering two extended X-ray sources on the same line of sight, and to  the presence of a diffuse emission bridge connecting the two lobes, we conclude that they most likely belong to the same extended X-ray object. 

To characterize this new object we extracted two spectra for its Eastern and Western lobes, and an overall spectrum encompassing the full source. The spectra were fitted with an absorbed power-law model, using \emph{Xspec}, and the instrumental and astrophysical background were estimated from the \emph{Off} region indicated in figure~\ref{fig:xmmregions} (top right panel). The results are presented in table \ref{tab:specresults}. Both lobes have a hard spectral index ($\Gamma = 1.6-1.8$) and highly absorbed spectra (n$_{\text{h}} \approx3 \times 10^{22}\,$cm$^{-2}$), confirming that they likely  belong to the same physical object.
The spectrum from the entire diffuse structure has an index of $1.99\pm0.45$, slightly softer than the average index of the two independent substructures. This is likely due to the fact that the box used for full source region also encompasses the bridge between the Eastern and Western lobes (see figure~\ref{fig:xmmregions}), whose  X-ray spectrum must be softer.

\banana is clearly detected in the \mbox{$2-10\,$keV} band (see figure~\ref{fig1}) but not at lower energies (see the $0.5-2\,$keV image in figure~\ref{fig:lowen}), due to the high absorption along the line of sight.  
Instead, in the \mbox{$0.5-2\,$keV} band, a point source appears to the South of \banana (see figure \ref{fig:diffuse_low_energy}). This source has an infrared counterpart (\mbox{GLIMPSE G344.0085-001748}) and a Gaia counterpart \citep[Source id 5966304379880060544 in Gaia DR3][]{gaia3}. 
The  parallax of the Gaia source is of 5.763$\,\pm\,$0.045$\,$mas, equivalent to a distance of about 170$\,$pc. This nearby distance rules out an association of the optical source with \banana whose high absorption indicates a larger distance (see section~\ref{disc_banana}).

As a conclusion, the newly discovered X-ray object \banana has an  extension of about $3'$ (along its longer axis) with two substructures likely belonging to same object, a high absorption and a hard spectral index.
The possible association of this source with \jsevA is discussed  in section \ref{disc_banana}.

\subsection{Spectral analysis of the \jsevA region}\label{ul}

Due to the absence of a clear X-ray signal,    we derived  upper limits on the diffuse X-ray emission in the \jsevA region (section~\ref{diffX}). Under a one-zone leptonic  hypothesis,  we then obtained   upper limits on the average magnetic field  by jointly fitting the X-ray (\emph{XMM-Newton}) and \gr (H.E.S.S.) data (section~\ref{aveB}). The multi-wavelength X-ray and \gr data  modeling was performed using the custom Gammapy-based scripts described in Appendix~\ref{gammapyX}. As a crosscheck, we repeated the analysis presented in section~\ref{diffX} using   the standard \emph{Sherpa} X-ray software~\citep{sherpa1, sherpa2}, obtaining consistent results (see figure~\ref{fig:validation}).

Obtaining a reliable upper-limit on an extended region is a challenging task since it requires an accurate description of the instrumental and astrophysical backgrounds in the \XMM field of view. 
As the instrumental background can strongly vary across the  field of view\footnote{\url{https://xmm-tools.cosmos.esa.int/external/xmm_user_support/documentation/uhb/epicintbkgd.html}}, it could not be estimated in a reliable way from different regions in the camera. Therefore we used a \emph{double subtraction technique}~\citep[see e.g.][]{2009A&A...505..157A,2013A&A...551A...7A} with a $0.06\ddeg$ circle (corresponding to the $1\sigma$ size of \jsevA)  as \emph{On} region and a nearby \emph{Off} region of the same size (see figure~\ref{fig:xmmregions}).
In each region the instrumental background was estimated locally from archival filter wheel closed (FWC) observations\footnote{\url{https://www.cosmos.esa.int/web/xmm-newton/filter-closed}}.
The \emph{Off} region was used to estimate the astrophysical background which we modeled as a
\texttt{TbAbs}-absorbed power law model.
In the \emph{On} region a model was added on top of the astrophysical background, in order to look for a possible X-ray signal in the \jsevA region. Namely, we tested an absorbed power law (section~\ref{diffX})   and an  absorbed electron synchrotron model (section~\ref{aveB}). 
The spectral parameters of the astrophysical background estimated from the \emph{Off} region were jointly fitted with the source model in the \emph{On} region. 
Due to strong spatial variation of the instrumental spectral lines  in the PN camera and despite our double subtraction method, the reliability of the PN instrumental background was not satisfactory for such an extended region and we therefore limited this analysis to MOS1 and MOS2 data.

\subsubsection{Upper limits on the diffuse X-ray emission }\label{diffX}
\begin{figure}
     \centering
         \includegraphics[width=\linewidth]{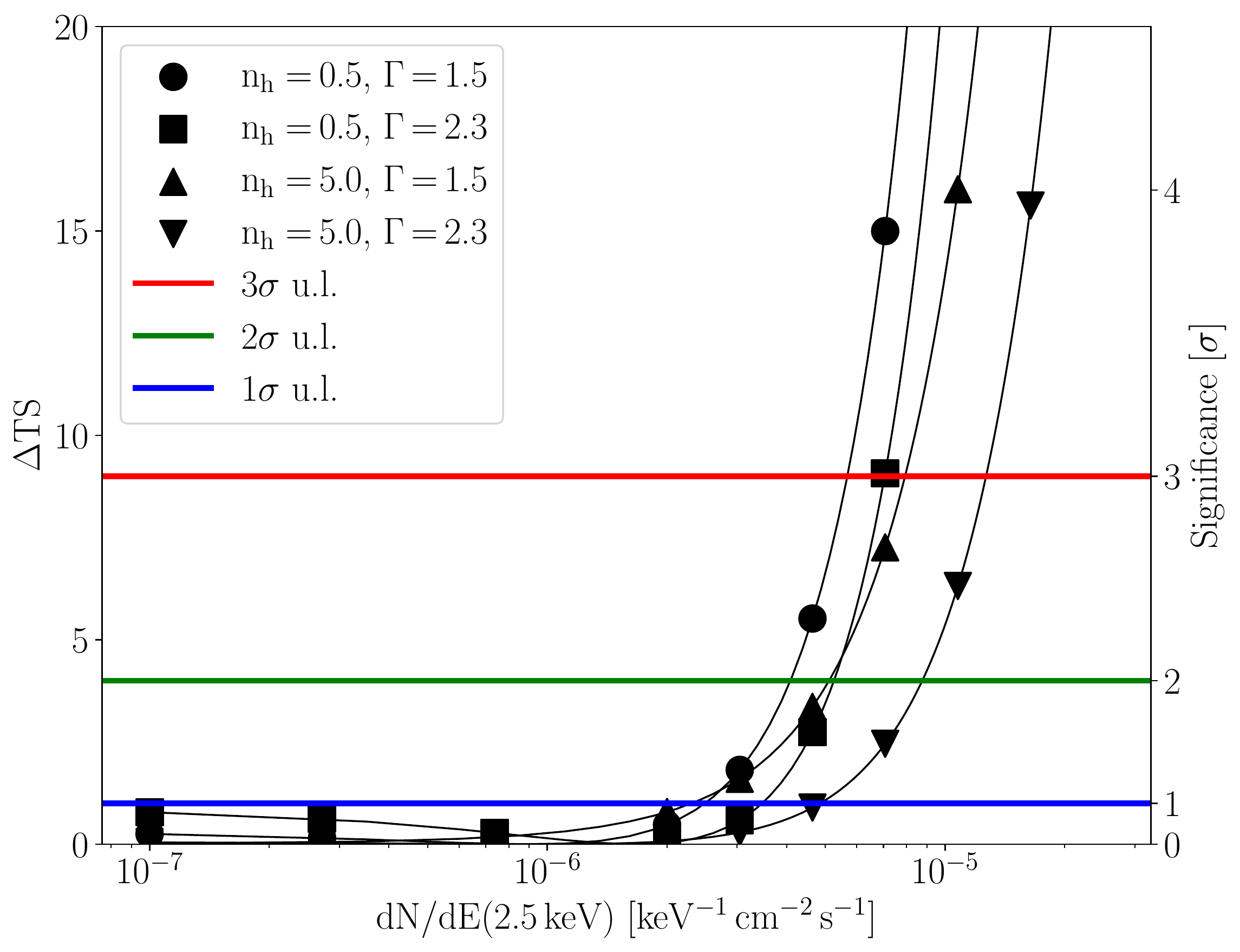}
        \caption{Likelihood profiles  for the de-absorbed differential flux at $2.5\,$keV of a putative X-ray signal inside the $0.06\ddeg$-radius \jsevA region, for different values of assumed absorption column density (n$_\text{h}$) and spectral index ($\Gamma$).}
        \label{uls_gamma}
\end{figure}

We  explored the  multi-dimensional likelihood surface by  varying the flux normalization of the additional  signal model (a \texttt{TbAbs}-absorbed power law) and simultaneously marginalized over all the free background nuisance parameters. This allowed us to derive likelihood profiles for the de-absorbed X-ray signal flux in the \jsevA region, for different assumptions on its spectral index and absorption column density (see figure~\ref{uls_gamma}). The assumed values for n$_\text{h}$ and $\Gamma$ correspond to extreme cases that are expected to enclose the real ones. We then computed the $(3, 2, 1)\,\sigma$   upper limits on the de-absorbed energy flux by integrating the corresponding spectra   between 2 and 10$\,$keV:
\begin{equation}\label{eflux}
    F(2-10\,\text{keV}) = \int_{2\,\text{keV}}^{10\,\text{keV}}E\frac{dN}{dE}dE\,.
\end{equation}
The upper limits obtained   are reported in table~\ref{FUL}. Each combination of assumed n$_\text{h}$ and $\Gamma$ generates   upper limits  with relative differences within $\approx30\%$. If we conservatively consider the highest upper limit for each   combination, the corresponding $3\sigma$ ($2\sigma$, $1\sigma$) constraints on the diffuse X-ray emission in the \jsevA region are:
\begin{equation}
    F(2-10\,\text{keV})<8.1\, (5.4,\,3.3)\,\times10^{-5}\,\text{keV}\,\CMminustwo\SECminusone.
\end{equation}

\subsubsection{Upper limits on the average  magnetic field}\label{aveB}
In order to constrain  the average  magnetic field inside the $0.06\ddeg$-radius \jsevA region,  we performed a multi-wavelength modeling of the X-ray and TeV \gr emission   using the new set of scripts described in Appendix~\ref{gammapyX}.  To describe  a hypothetical X-ray signal in the \jsevA region,  we   defined an electron  synchrotron  model  with \texttt{TbAbs} absorption (obtained as a combination of the corresponding \emph{Naima} and \emph{Xspec} models within a common \gp framework). The synchrotron model depends on the ambient magnetic field value B and on the parameters of the assumed $e^\pm$ distribution, here an exponential-cutoff power law (ECPL):
\begin{equation}\label{elec}
               \frac{dN_e}{dE_e}(E_e) = A_e \left(\frac{E_e}{E_e^*}\right)^{-\Gamma_e} \text{exp}\left[-\left(\frac{E_e}{E_e^\text{cut}}\right)\right]\,.
\end{equation}
The same electron population was used also to predict the corresponding VHE \gr emission  due to inverse-Compton up-scattering of cosmic microwave background (CMB) photons. The choice of such a minimal one-zone leptonic model was  justified  by the absence of a clear X-ray signal, which did not motivate more complex assumptions. 
%In order to reduce the high dimensionality of the problem, we explored the   likelihood surface along the axes of constant electron cutoff energy  and  absorption density for the  synchrotron model. 
%\footnote{The  number of degrees of freedom was 8:  three for the astrophysical X-ray background (described by equation~\ref{abspl}),   three for the electron population (equation~\ref{elec}), one for the absorption of the synchrotron model and plus the ambient magnetic field B.}

\begin{figure*}
     \centering
     \begin{subfigure}[b]{0.49\textwidth}
         \centering
         \includegraphics[width=\textwidth]{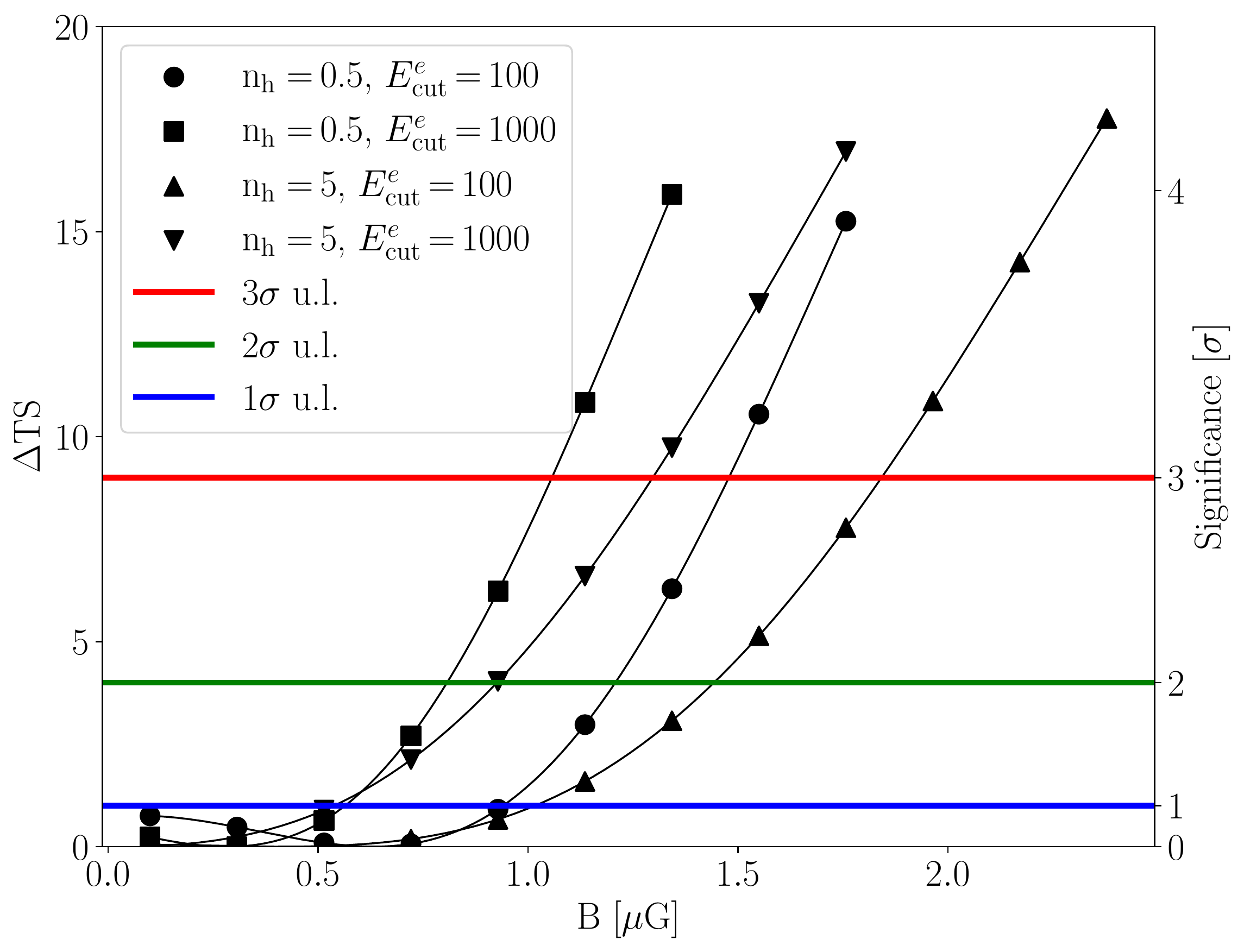}
     \end{subfigure}
     \hfill
     \begin{subfigure}[b]{0.49\textwidth}
         \centering
         \includegraphics[width=\textwidth]{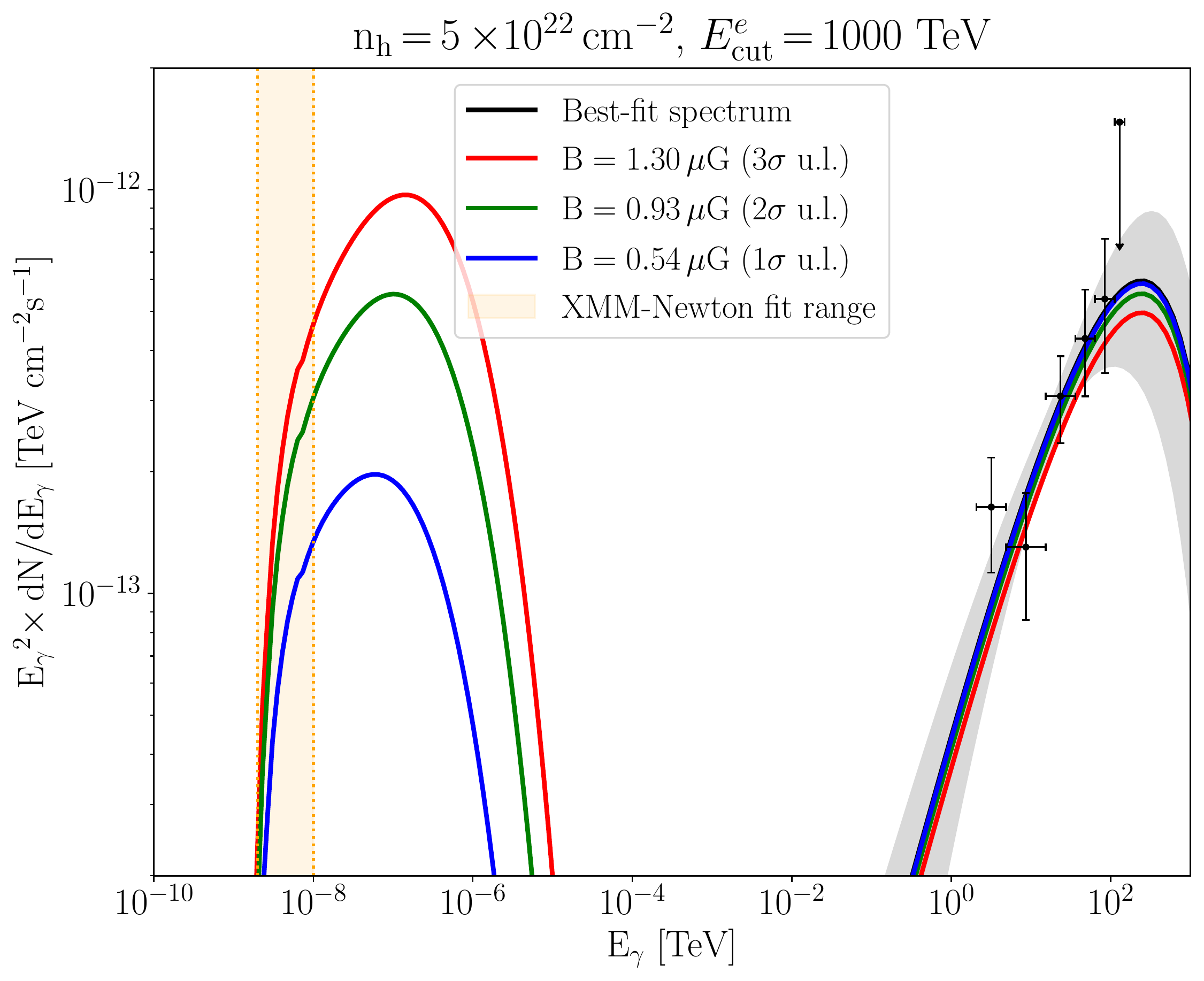}
     \end{subfigure}
        \caption{\emph{Left:} Likelihood profiles  for the average magnetic field inside the $0.06\ddeg$-radius \jsevA region, assuming a one-zone leptonic model, for different values of assumed absorption column density (n$_\text{h}$) and electron cutoff energy ($E_\text{cut}^e$). \emph{Right:} Synchrotron and inverse-Compton spectra corresponding to the $3\sigma$ (red), $2\sigma$ (green) and $1\sigma$ (blue) magnetic field upper limits in the \jsevA region, together with the best-fit inverse-Compton model for the H.E.S.S.\ SED (black). For this illustrative plot, a n$_\text{h}=5\times10^{22}\CMminustwo$, $E_\text{cut}^e=1\PEV$ and minimum electron energy of 100$\,$GeV were assumed. An electron spectral index of $\Gamma_e=1.48\pm0.27$, obtained from a fit of the H.E.S.S.\ data points alone, was also assumed.}
        \label{BBB}
\end{figure*}

We adjusted the parameters of the electron distribution (equation~\ref{elec}) to the publicly available TeV SED of \mbox{HESS J1702-420A}~\citepalias{Giunti}, that was re-scaled to account for the finite emission fraction contained inside the   $0.06\ddeg$-radius spectral extraction region\footnote{Considering that the $1\sigma$ radius of a two-dimensional Gaussian encloses $1-e^{-1/2}\approx0.3935$ of its integral, we multiplied the SED provided by~\citetalias{Giunti} by a factor of 0.39.}. We then froze the value of the electron spectral index, and fitted all the remaining model parameters simultaneously to the \XMM data spectra and H.E.S.S.\ SED data points. This allowed us to compute the likelihood profiles for the magnetic field   that are shown in  figure~\ref{BBB} (left panel). The resulting magnetic field upper limits are reported in table~\ref{BUL}. As an example, the right panel of figure~\ref{BBB} shows the inverse-Compton and synchrotron models corresponding to the 1, 2 and $3\sigma$ upper limits on B, for a particular choice of (fixed) electron cutoff and absorption column density. %A consistency check between the upper limits derived in sections~\ref{diffX} and~\ref{aveB} is provided in Appendix~\ref{consistency}.

Independently of the assumed values  for the nuisance parameters, the magnetic field is constrained to be lower than    $1.45\,\mu$G at $2\sigma$, approximately corresponding to a 95.5\% confidence level. This value is   low, compared to the average Galactic magnetic field of $\approx3\,\mu G$, a fact that challenges the leptonic emission hypothesis for \jsevAvirg at least under a simple one-zone scenario. %, and strengthens its  figure~\ref{BBB} (left panel) and table~\ref{}, the upper limits are pulled toward more conservative values when an extremely low electron cutoff of $E_e^\text{cut}=100\,$TeV is chosen. Since \citetalias{Giunti} had estimated an electron cutoff lower limit around  $E_e^\text{cut}>=??\,$TeV for \jsevAvirg we argue that the upper limit estimate driven by the $E_e^\text{cut}=100\,$TeV is likely too conservative and find it more appropriate to quote a $2\sigma$ confidence level lower limit of $B\lesssim1\,\mu$G.

\section{Discussion}\label{discussion}
This section presents  the implications of the new \XMM  observation on the   nature of \jsevAvirg focusing   on its possible association with \mbox{\Suzaku src B} (section~\ref{disc_suzB}) and \banana (section~\ref{disc_banana}).

\subsection{Is \jsevA associated with \Suzaku src B?}\label{disc_suzB}
\begin{figure*}
     \centering
     \begin{subfigure}[b]{0.49\textwidth}
         \centering
         \includegraphics[width=\textwidth]{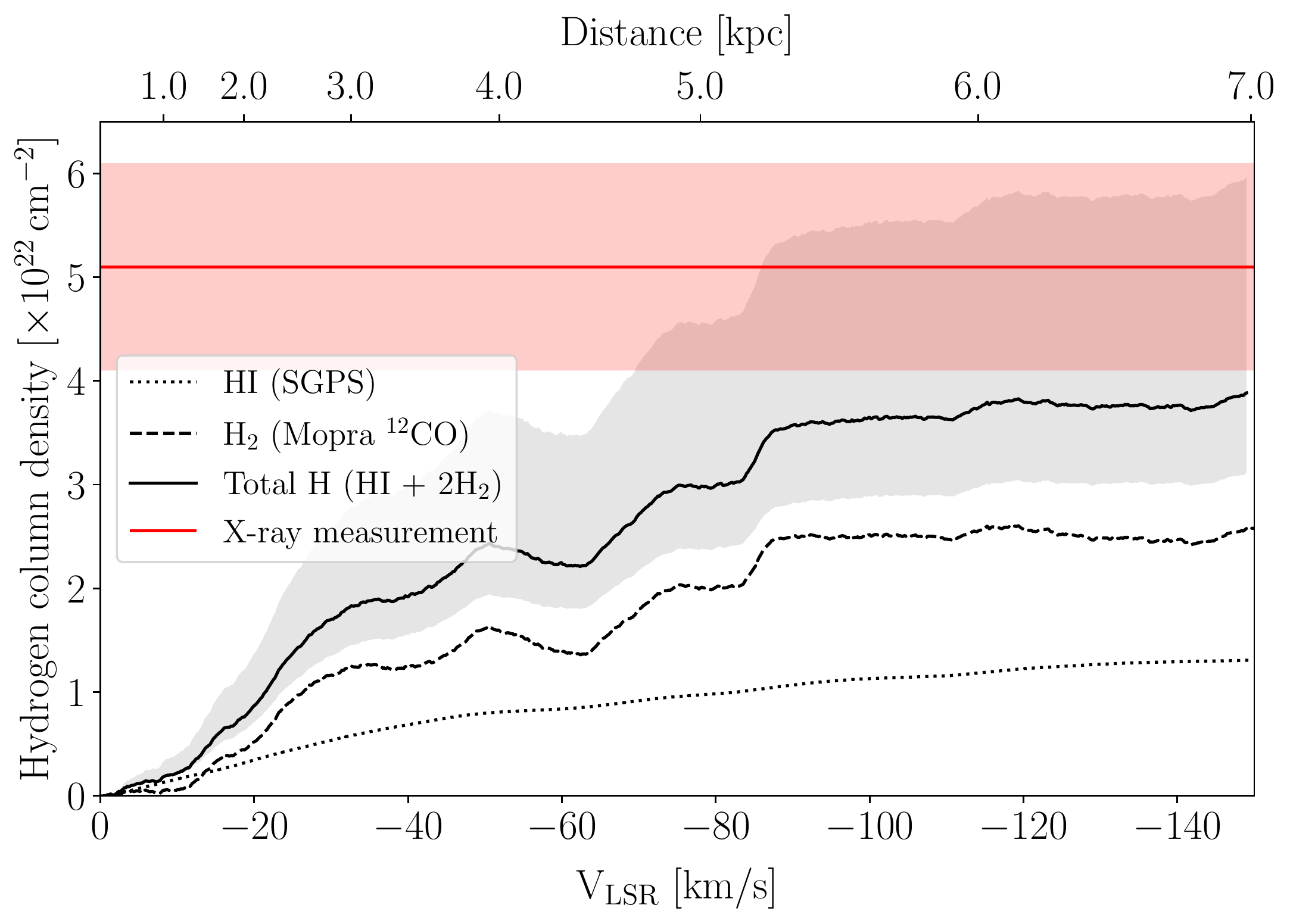}
     \end{subfigure}
     \hfill
     \begin{subfigure}[b]{0.49\textwidth}
         \centering
         \includegraphics[width=\textwidth]{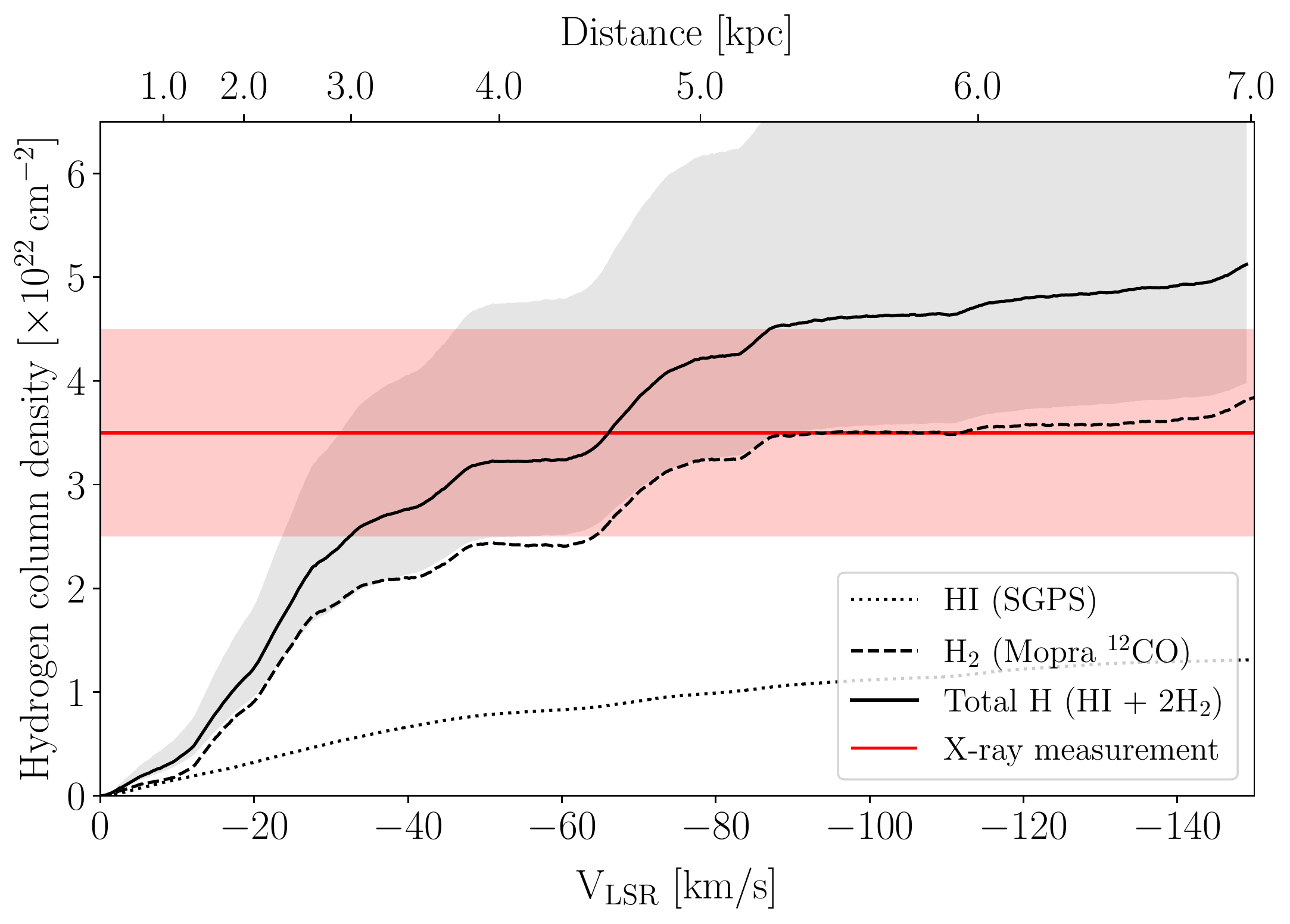}
     \end{subfigure}
    \caption{Cumulative hydrogen column density in the direction of \Suzaku src B (\emph{left}) and \banana (\emph{right}).  The shaded gray region indicates the total level of systematic uncertainties, due the assumption of an average conversion factor  \mbox{$X_\text{CO} = 1.5\times10^{20}\,  \text{cm}^{-2}\, (\text{K}\,\text{km}\,\text{s}^{-1})^{-1}$~\citep{2004A&A...422L..47S}} which impacts the H$_2$ measurement up to   30\%, and an optically thin medium which can lead to underestimate  the HI content up to factor of 2~\citep{lau}. The red band indicates the modeled column density required by the X-ray observations, with statistical errors only (see table~\ref{tab:specresults}).  }
    \label{fig:suzaku_abs}
\end{figure*}
\citetalias{Giunti}  have suggested that the unidentified source \mbox{\Suzaku src B} could be a powerful isolated  pulsar responsible for the VHE \gr emission of \jsevA via a leptonic (inverse-Compton) emission scenario.
This idea is now challenged by the non-detection of an extended X-ray synchrotron nebula around  \mbox{\Suzaku src B}, which appears to be point-like in the \XMM data (see figure~\ref{fig:XMM_spec}, bottom left panel). This argument is however not conclusive, since evolved $\gamma$-ray bright PWNe can have suppressed X-ray emission due to the complete cooling of their most energetic electrons~\citep{Acero:2017rgv}. However the spectral analysis presented in section~\ref{suzakuB} can be used  to discuss a possible pulsar-PWN  association between \mbox{\Suzaku src B} and \jsevAdot

 An X-ray absorption column density as high as \mbox{n$_\text{h}=(5.1\pm 1.1)\times 10^{-22}\,\text{cm}^{-2}$}  implies that \mbox{\Suzaku src B} is located further away than  $7\,$kpc. This is shown in figure~\ref{fig:suzaku_abs} (left panel), which compares  the absorption level required by the X-ray measurement with the total cumulative hydrogen column density measured in the Mopra~\citep{mopra} and SGPS~\citep{sgps} radio surveys, averaged over a $15''$ radius circle centered on \Suzaku src B. The total gas density   integrated up to a distance of $7\,$kpc   never reaches that derived from the X-ray spectral fit.  The gas velocities were converted to distance measures using the Galaxy model presented in~\citet{Vall_e_2016}, and assuming  that all the gas up to $-150\,\text{km}\,\text{s}^{-1}$ is located at the near kinematic distance.  
The H.E.S.S.\ experiment is capable of detecting point-like sources around  $7\,$kpc away only if they are  intrinsically bright ($\gtrsim10^{34}\,$erg$\,\text{s}^{-1}$)~\citep{hgps}, with a required brightness for extended sources that is even higher. Therefore, regardless of the real nature of \Suzaku src B, the high X-ray absorption measurement by itself does not suggest an   association with the extended source \jsevAdot %Still, we note that when the systematic uncertainties on the radio measurements (gray shaded band) and the statistical error on the X-ray fit  are taken into account, the distance lower limit lowers down to $5\,$kpc. So, even this argument is not by itself sufficient to definitely exclude an association with \jsevAdot

  Perhaps the most convincing reason to reject a classification of \Suzaku src B as an isolated pulsar comes from the detection of an Fe spectral line in its spectrum,  which implies instead an accreting compact object such as a cataclysmic variable (CV) system with thermal emission from the   disk. This hypothesis is supported by the measured energy of the Fe line, $(6.62\pm 0.06) \,$keV, which is compatible with a mixture of the $6.4\,$keV and $6.7\,$keV Fe lines that are observed in other magnetized  CVs~\citep{2016ApJ...818..136X}.  The hard  measured X-ray spectrum is also in line with the expectation for magnetized CVs~\citep{2012A&A...542A..22B, 10.1111/j.1365-2966.2009.15826.x}.
  Additional  support to this hypothesis is provided by the evidence of spectral variability deriving from a  comparison between the new flux measurement in the $2-10\,$keV band   and the one carried out in 2008 with \Suzaku by  \citetalias{fujinaga},
  the former being significantly higher (by a factor of 10) than the latter. This fact, also illustrated in figure~\ref{variab}, indicates  a  strong flux increase between 2008 and 2021, which is not an expected feature for an isolated pulsar. 
  %We note that the \XMM and the \Suzaku flux estimates are not directly comparable, since  based on very different spectral assumptions, since \citetalias{fujinaga} had assumed a fixed index of 2.1 in the absence of sufficient counts to carry out an X-ray spectral measurement. However, we estimated that assuming the 2021 \XMM spectral model (see Table \ref{tab:specresults}), the WebPIMMS-predicted number of \Suzaku counts that should have been detected with a 216$\,$ks \Suzaku is 1400 counts compared to the 228 counts reported in the paper confirming a strong spectral variation between 2008 and 2021.
  %pointing remains $\approx3.6$ times higher than the one reported in  \citetalias{fujinaga}. This could be either due to the high level of systematics affecting the \Suzaku measurement, as the fact that the source was possibly not  entirely contained in the field of view, or point to  a real spectral variability that is typical of CVs. 
The existence of a possible  infrared counterpart to \Suzaku src B, called \mbox{GLIMPSE G344.0901-00.1658}, also points in the same direction. 

In conclusion, our deep \XMM  observation provides sufficient evidence to  affirm that, regardless of its real nature,  \Suzaku src B is not an isolated pulsar associated with \jsevAdot Further studies will be needed to investigate its possible classification as a magnetized CV.

%\subsection{Comparison of X-ray and VHE gamma-ray fluxes}\label{tev_x_ratio}
%Discuss the TeV to X ratio, as done in Fujinaga 2011. It is quite high but not super high if we integrate between 2 and 10 TeV, much less than estimated by fujinaga. But of course this is partly due to the spectral hardness 
%\begin{equation}\label{ratio}
 %              \frac{F_{1-10\,\text{TeV}}^\text{H.E.S.S.}}{F_{2-10\,\text{keV}}^\text{XMM}} \gtrsim 3.2\, (4.8,\, 7.8)\,.
%\end{equation}

\subsection{Is \jsevA associated with \mbox{XMMU J170147.3-421407}?}\label{disc_banana}
\begin{figure}
     \centering
         \includegraphics[width=\linewidth]{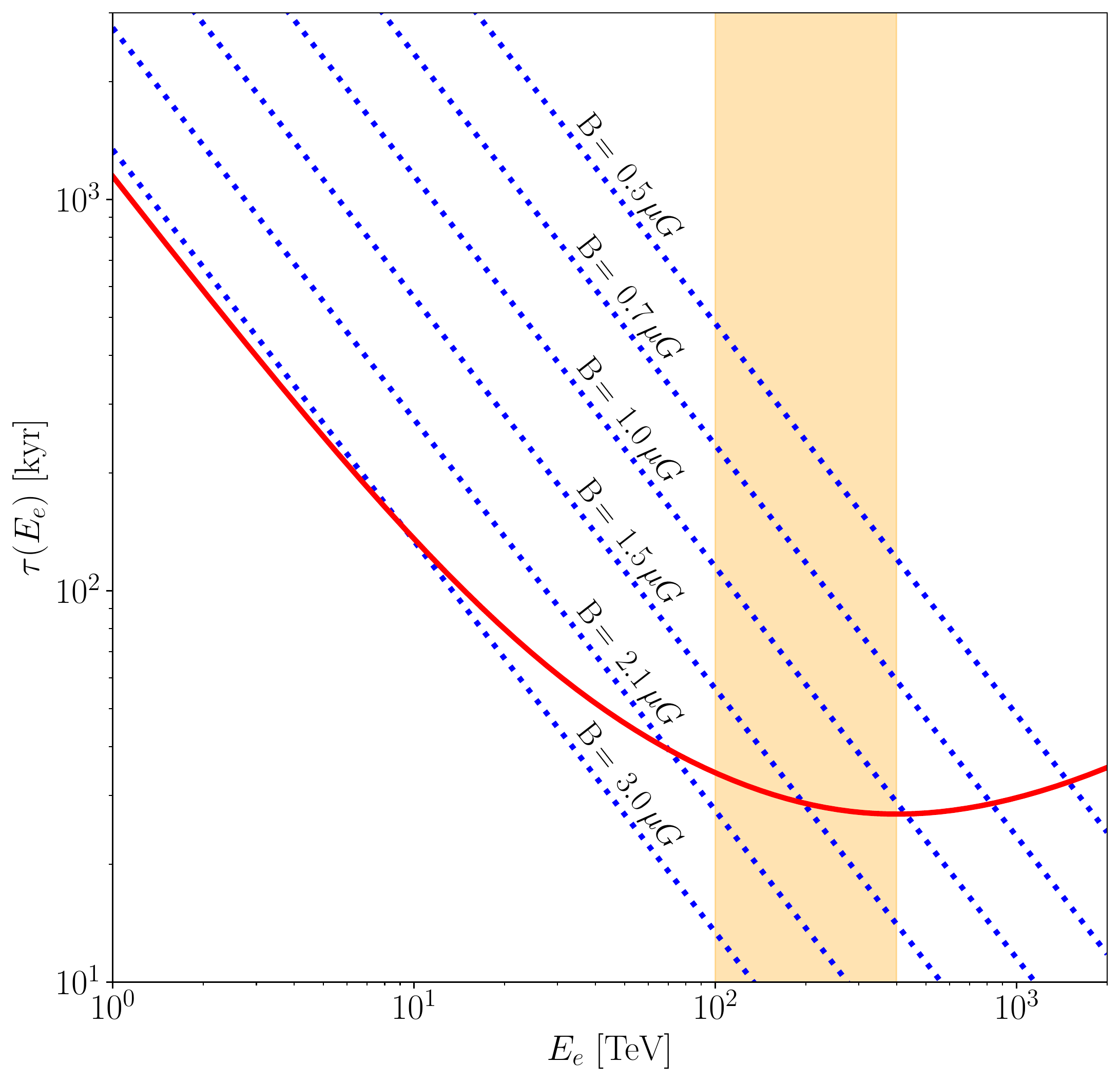}
        \caption{Electron cooling time due to the inverse-Compton (red) and synchrotron (blue) processes. The yellow band indicates electron energy in the range $200-400\,$TeV, as needed to create the TeV emission of \jsevA (see the main text).}
        \label{tau}
\end{figure}
After excluding an association between \jsevA and \Suzaku src B (section~\ref{suzakuB}) and estimating an unrealistically low magnetic field in the \jsevA region, under  a single-zone leptonic hypothesis (section~\ref{BUL}), any simple leptonic scenario for the origin of \jsevA appears to be   constrained. An open question is whether \banana (section~\ref{banana}) can be a PWN  associated with \jsevAdot Indeed, the chance of discovering an extended X-ray source with hard spectral index nearby  a hard \gr source like \jsevA  by simple chance is not expected to be high. 

To be a PWN associated with \jsevAvirg \banana  must firstly be a Galactic object, rather than an extra-Galactic one such as an active Galactic nucleus. We therefore compared the absorption measure obtained from the X-ray fit to the cumulative hydrogen density on the line of sight, with the same method presented in section~\ref{disc_suzB}, to investigate whether \banana can be located inside the Milky way or not. As shown in figure~\ref{fig:suzaku_abs} (right panel), the best estimate that we can provide for the distance of \banana (given by the intersection of the solid red and black lines)   is $\approx4.7\,$kpc. Despite the large statistical and systematic uncertainties, suggesting  that the actual source distance could be  in the $\approx2-8\,$kpc range, its interpretation as a Galactic object appears favored. 

Another element that must be taken into account is the   strongly asymmetric   morphology of \bananav characterized by two distinct lobes with hard X-ray index connected by a softer bridge. Despite being immediately suggestive of a microquasar interpretation, this peculiar shape    is   not incompatible with a PWN interpretation either. Indeed, previously reported cases of energetic PWNe show very similar structures, with double-jet morphologies characterized by extremely hard X-ray spectral indices~\citep{2017A&A...597A..75M}.

The remaining piece of evidence to be discussed is the $\approx0.12\ddeg$ offset between the centroids of \jsevA and \bananav  which complicates their association since the X-ray emission from other known PWNe usually coincides with their   hardest TeV emission peak~\citep{j1825}. Still, it remains possible that \banana is   a PWN  powered by a yet undetected pulsar that has drifted away from the center of \mbox{HESS J1702-420A} due to a large proper motion, similarly e.g.\ to the Mouse and Geminga   bow-shock nebulae~\citep{mouse, geminga}.  
In order to  explore this hypothesis, we computed the cooling time of the  electrons that may be responsible for the TeV emission of \jsevAvirg which is an upper limit on their age,  and used it to set a lower limit on the putative pulsar's runaway speed. 
  The electron cooling time via inverse-Compton emission on CMB photons can be estimated using the analytic approximation presented in equation 40 of~\citet{khangulyan_IC}, while the synchrotron cooling time was computed following~\citet{RevModPhys.42.237}. 
 Figure~\ref{tau}  compares the inverse-Compton and synchrotron cooling times as a function of the electron energy, for different values of magnetic field B.  The energy of the putative electrons producing the $50-100\TEV$  \gr emission of \jsevA via inverse Compton was estimated to be in the \mbox{$100-400\TEV$} range (indicated by the yellow band in figure~\ref{tau}), using the equation 1 from~\citet{doi:10.1126/science.abg5137}. If we assume for example that $B\lesssim1\mu$G, electrons in the $100-400\TEV$  energy range  are predominantly cooled within a time scale of  $\approx30\,$kyr (inverse-Compton dominated losses). This means that the putative electron population responsible for the VHE emission of \jsevA must have an age $t_e\leq30\,$kyr. Assuming that \jsevA and \banana are connected,  such a limit on the electron age implies that the X-ray source must be running away at a speed $v\gtrsim d\tan(0.12\ddeg)/t_e$, where $d$ is the distance from Earth of \jsevAdot For a distance $d$ of 2 (8) kpc, the resulting lower limit on the velocity of \banana is of $\approx$140 (550) km$\,$s$^{-1}$, which is entirely compatible with the typical values for young runaway pulsars~\citep{Hobbs_2005, pulsars}.  
 
 In conclusion, we found that there is at least one scenario (high-speed runaway PWN in an inverse-Compton dominated  cooling environment)  in which an association between \jsevA and the \banana is possible.
 We note however that, by requiring a $B\lesssim1\mu$G, this is not an obvious interpretation  that will need to be further investigated. In particular, observations  with better angular resolution (both in the X-ray and radio bands)  will allow to search for a point-like source  inside the diffuse structure of \bananav whose discovery would potentially confirm its  PWN nature and strengthen its association with \jsevAdot
 
 \section{Conclusions}\label{conclusions}
This paper describes a new deep X-ray observation of \jsevAvirg the hard VHE \gr emission region inside \jsevvirg carried out with \emph{XMM-Newton}.
For a number of other initially  unidentified H.E.S.S.\ sources,   the discovery of     X-ray counterparts has represented the smoking gun pointing to a PWN classification, which implies a leptonic origin of their bulk TeV emission~\citep{j1303, j1825, 2022arXiv220403185B}. Here  we demonstrate that the case of \jsevA is different: 
this object is completely undetected in the X-ray range, a fact that challenges a simple leptonic interpretation of its TeV emission.

We first studied \Suzaku src B, the most obvious candidate counterpart of \jsevAvirg and concluded that it is not an isolated pulsar powering a \gr bright PWN. This conclusion  derived from a number of clues such as its point-like X-ray morphology, its large distance from Earth, its  variable spectrum with a clear Fe line and the existence of a possible infrared counterpart. All these  elements concur instead in making of \Suzaku src B a new  cataclysmic variable source candidate.

In the absence of a clear X-ray  detection, we derived   upper limits on the diffuse de-absorbed X-ray flux inside a $0.06\ddeg$-radius region corresponding to the $1\sigma$ contour of \mbox{HESS J1702-420A}: $F(2-10\,\text{keV})\lesssim5.4\,\times10^{-5}\,\text{keV}\,\CMminustwo\SECminusone$ at $2\sigma$ ($\approx95.5\%$) confidence level. Assuming a one-zone leptonic model, in which the  same electron population is radiating both in the X-ray (synchrotron) and VHE \gr (inverse-Compton) bands, we derived a tight constraint on the magnetic field by jointly fitting the \XMM spectra and the H.E.S.S.\ spectral energy distribution (SED): $B\leq1.45\,\mu$G. This  argument   discourages a   leptonic interpretation for the \gr emission of \jsevAvirg at least  under a minimal one-zone hypothesis.

Finally, we report the discovery of a new  X-ray source with extended ($\approx3'$ along its longer axis)  morphology and hard spectral index ($1.99\pm0.45$), named \bananap Given its absorption measure (n$_{\text{h}} \approx3 \times 10^{22}\,$cm$^{-2}$), this unidentified object is most likely Galactic. Its peculiar shape, characterized by two distinct lobes connected by a diffuse bridge, is compatible with previously-reported cases of PWNe with double-jet morphologies. Its  $\approx0.12\ddeg$ offset from the center of \jsevA   does not prevent an association, provided that \banana is a PWN powered by a high-speed runaway pulsar ($v\gtrsim 140-550\,\text{km}\,\text{s}^{-1}$) and that the electrons inside   \jsevA   are primarily cooled down by inverse-Compton losses on the CMB field. In that case, a multi-zone leptonic scenario in which \banana is powered by freshly injected electrons and \jsevA by electrons with an age $\lesssim30\,$kyr could not be excluded. One way to probe it could be looking for a faint X-ray or radio tail between the X-ray source and \jsevAdot Still, we note that this hypothesis requires an extremely low magnetic field ($B\leq1\,\mu$G) in the \jsevA region, and therefore  does not appear to be straightforward.

In conclusion, the hard \gr object \jsevA remains unidentified, but the absence of a clear X-ray counterpart strongly challenges simple leptonic scenarios. The only remaining possibility for a leptonic interpretation of \jsevA appears to be an association with a nearby extended X-ray source with hard spectral index, that could be powered by a high-speed young pulsar. This possibility will need to be further investigated by means of X-ray or radio observations with better angular resolution. 
Finally, this work contains a multi-wavelength X-ray and \gr modeling based on a new set of scripts based on \emph{Gammapy}, \emph{Naima} and \emph{Xspec}~\citep{giunti_luca_2022_7092736}. These scripts  demonstrate the potential of the open-science approach  and  
may be  applied in the future to other analysis cases. 

\section*{Acknowledgements}
The authors acknowledge finantial support from Agence Nationale de la Recherche (grant ANR- 17-CE31-0014) and from LabEx UnivEarthS (ANR-10-LABX-0023 and ANR-18-IDEX-0001).
 This research made use of Astropy\footnote{\url{http://www.astropy.org}}~\citep{astropy:2013, astropy:2018},  Numpy~\citep{harris2020array}, iminuit~\citep{iminuit} and Matplotlib~\citep{Hunter:2007}.
 
\bibliographystyle{aa} % style aa.bst
\bibliography{biblio.bib} % your REFerences Yourfile.bib

\appendix

\section{New recipes for the multi-wavelength modeling of X-ray and \gr data}\label{gammapyX}
We have developed a set of \emph{Python} scripts dedicated to the multi-wavelength fit of X-ray and $\gamma$-ray data with physically motivated models,  based on the \emph{Gammapy}, \emph{Naima} and \emph{Xspec} packages.
They were  developed on \emph{GitHub}~\footnote{\href{https://github.com/registerrier/gammapy-ogip-spectra}{https://github.com/registerrier/gammapy-ogip-spectra}} in the context of this work, but they may be generally applied to any spectral analysis of X-ray data (alone or jointly with \gr spectral and/or 3D data). They are therefore  made available for download on the \emph{Zenodo} archive~\citep{giunti_luca_2022_7092736}, to encourage their re-use and improvement in an open-source and open-science spirit. 

The idea behind the analysis scripts   is to   create a framework in which the X-ray and \gr data are handled in a consistent way, thanks to the  \texttt{Dataset} class of \emph{Gammapy}, and fitted together using the model libraries provided by \emph{Xspec} and \emph{Naima}.
The main building blocks of the scripts  are the \texttt{StandardOGIPDataset} and the  \texttt{SherpaSpectralModel} classes. The \texttt{StandardOGIPDataset} is a special type of Gammapy \texttt{Dataset} dedicated to X-ray spectral data, implementing X-ray-specific features such as the energy \emph{grouping}.   The \texttt{SherpaSpectralModel} is instead a class that wraps the \emph{Xspec} models (contained in the \texttt{sherpa.astro.xspec} module) in such a way that they can be used for fitting with \emph{Gammapy}. Together with the  \texttt{NaimaSpectralModel} feature of \emph{Gammapy}, this allows to adjust the electron or proton distribution parameters   directly   to the measured X-ray and \gr data using the \emph{forward-folding}\footnote{This means performing a likelihood-based match between the measured and model-predicted data after taking into account the instrument response functions.} technique~\citep{piron}. For example, as demonstrated in section~\ref{aveB}, in this framework one can easily fit X-ray data with a combination of absorption and electron synchrotron models, eventually including also  black-bodies and spectral lines, while simultaneously modeling the inverse-Compton \gr due to the same electron distribution.

\section{Additional material}
\begin{figure*}
    \centering
    \includegraphics[width=\linewidth]{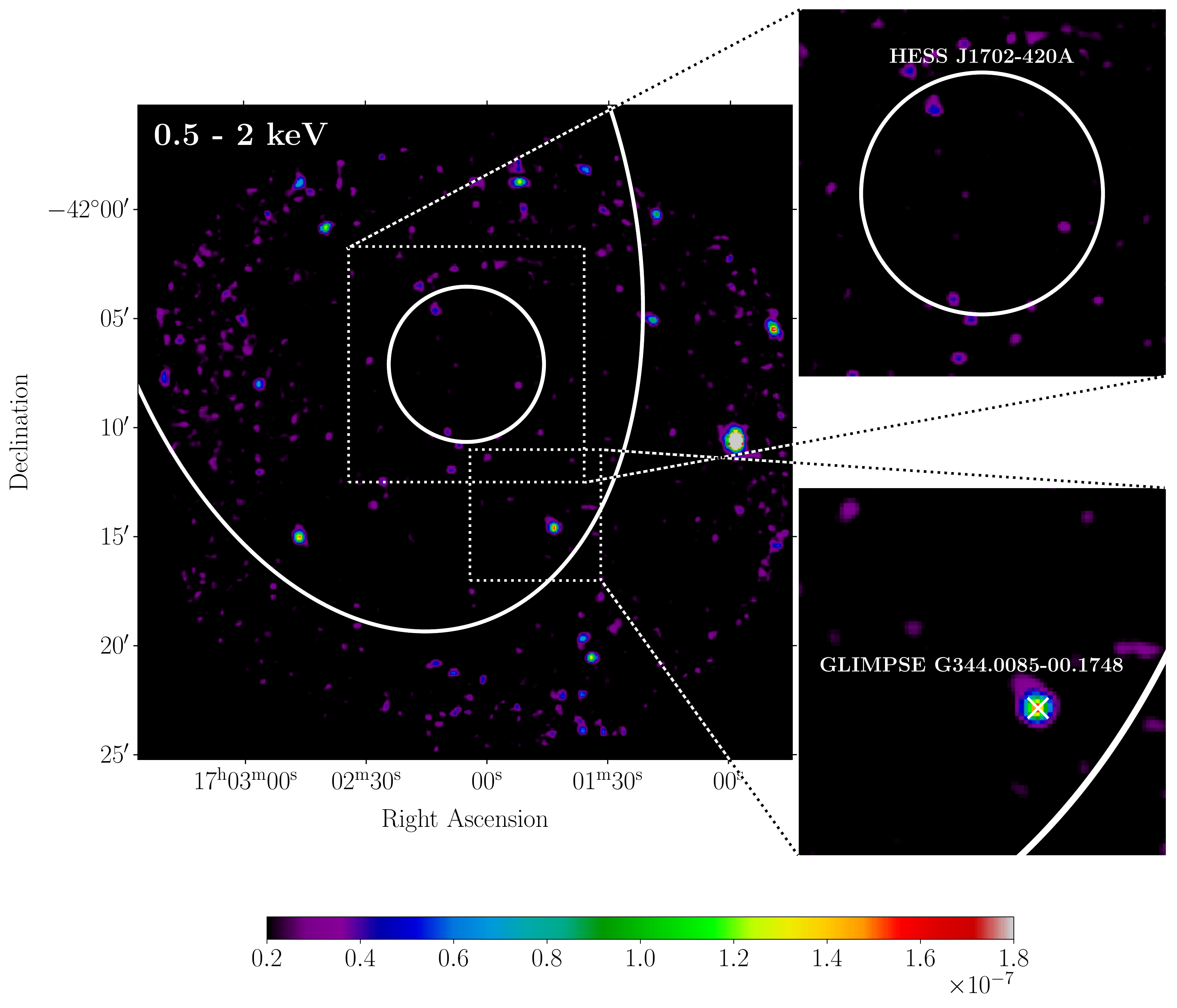}
    \caption{Same as figure~\ref{fig1}, but integrated in the $0.5-2\,$keV energy band.}
    \label{fig:lowen}
\end{figure*}

\begin{figure*}
\begin{subfigure}{.5\linewidth}
\centering
\includegraphics[width=1.08\linewidth]{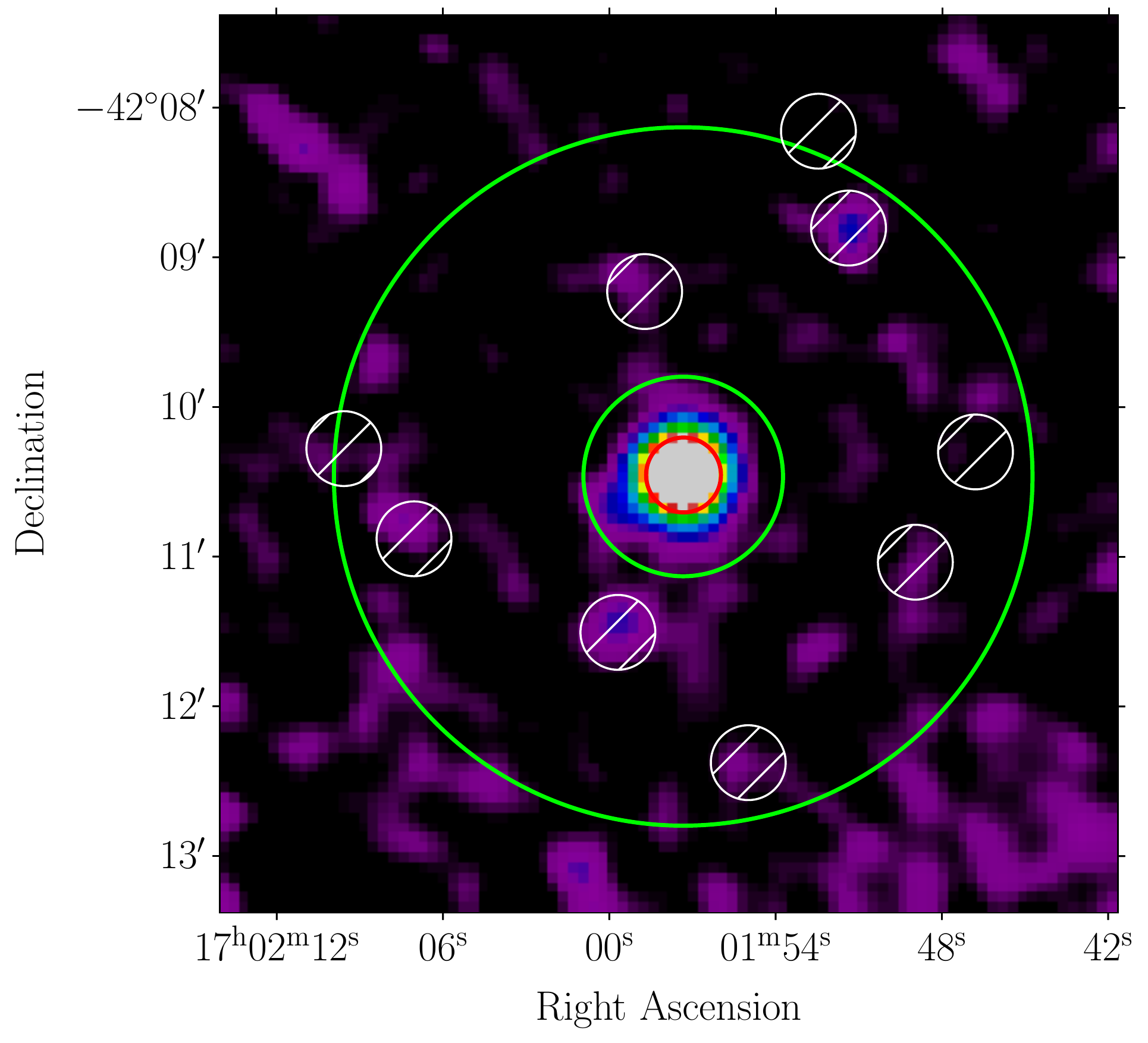}
\end{subfigure}%
\begin{subfigure}{.5\linewidth}
\centering
\includegraphics[width=0.8\linewidth]{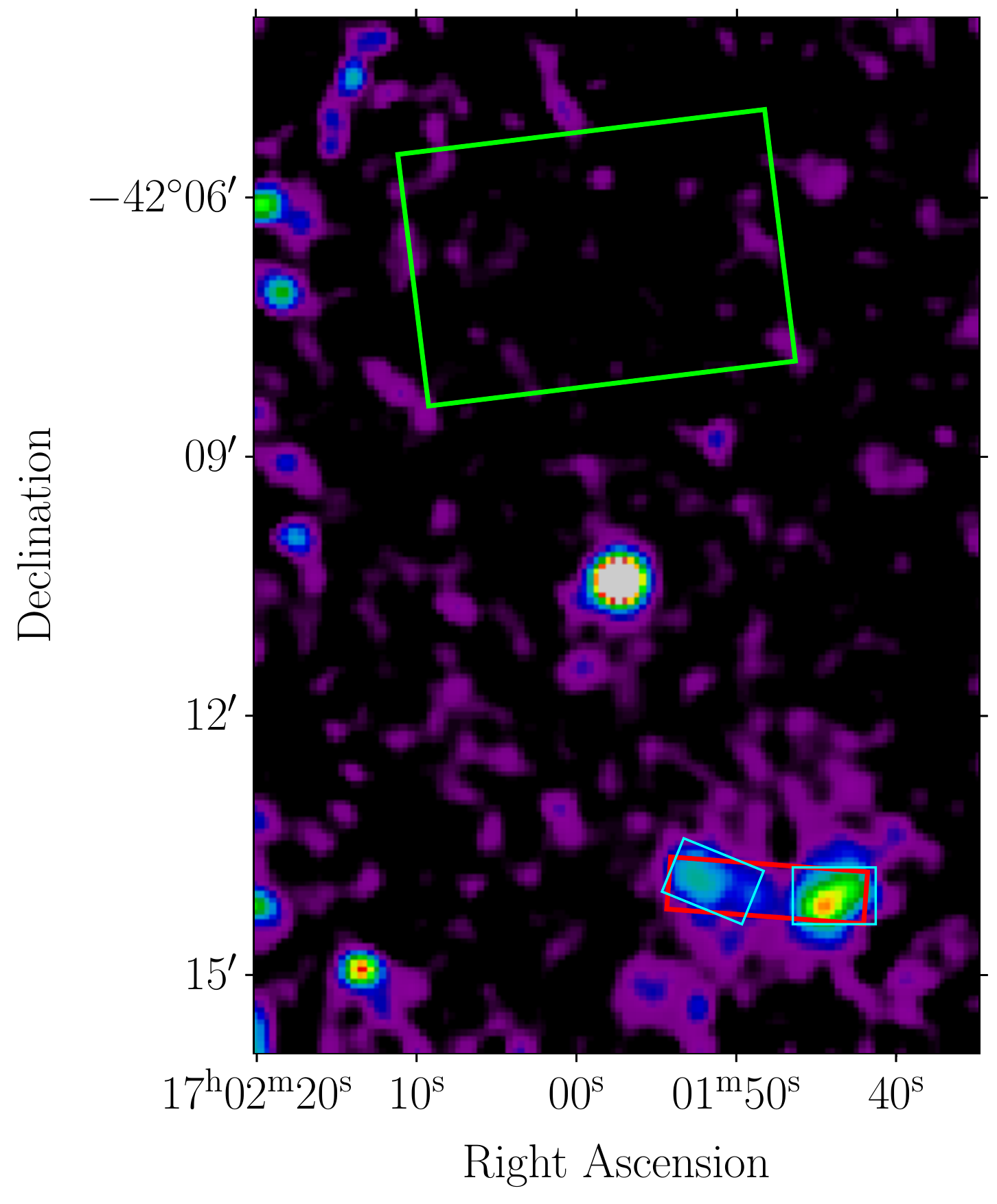}
\end{subfigure}\\[1ex]
\begin{subfigure}{\linewidth}
\centering
\includegraphics[width=0.7\linewidth]{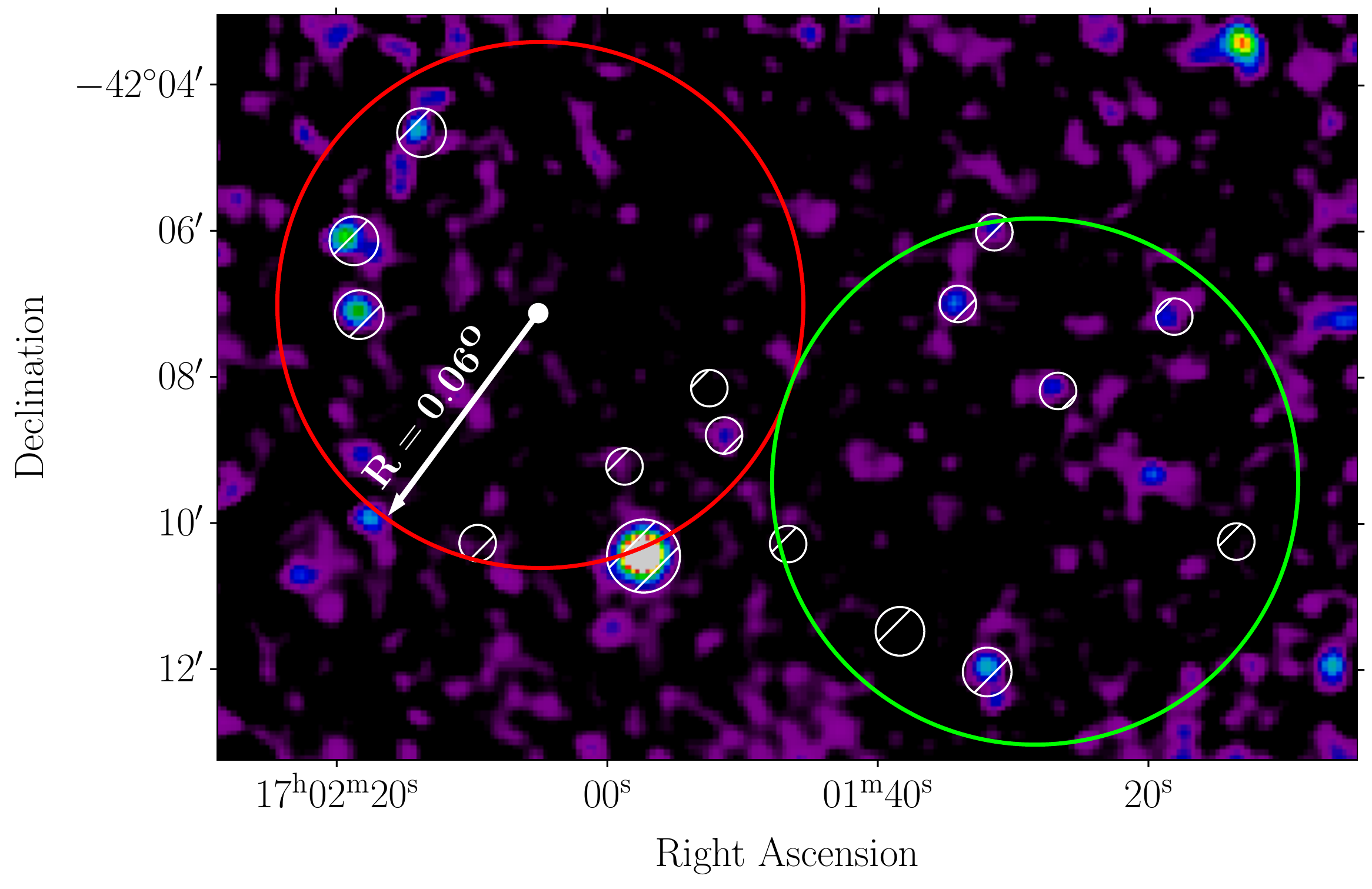}
\end{subfigure}
\caption{Spectral extraction regions for the three analyses described respectively in sections~\ref{suzakuB} (upper left panel),~\ref{banana} (upper right panel) and~\ref{ul} (lower panel). In all cases the \emph{On} region is shown in red, the background control region in green and the masked fluctuations in white. In the upper right panel, the \emph{On} regions used for the spectral analysis of the Eastern and Western lobes of \banana are shown in cyan. All images show the \XMM data (sum of PN and MOS cameras) in the $2-10\,$keV band.}
\label{fig:xmmregions}
\end{figure*}
\begin{figure}
    \centering
    \includegraphics[width=\linewidth]{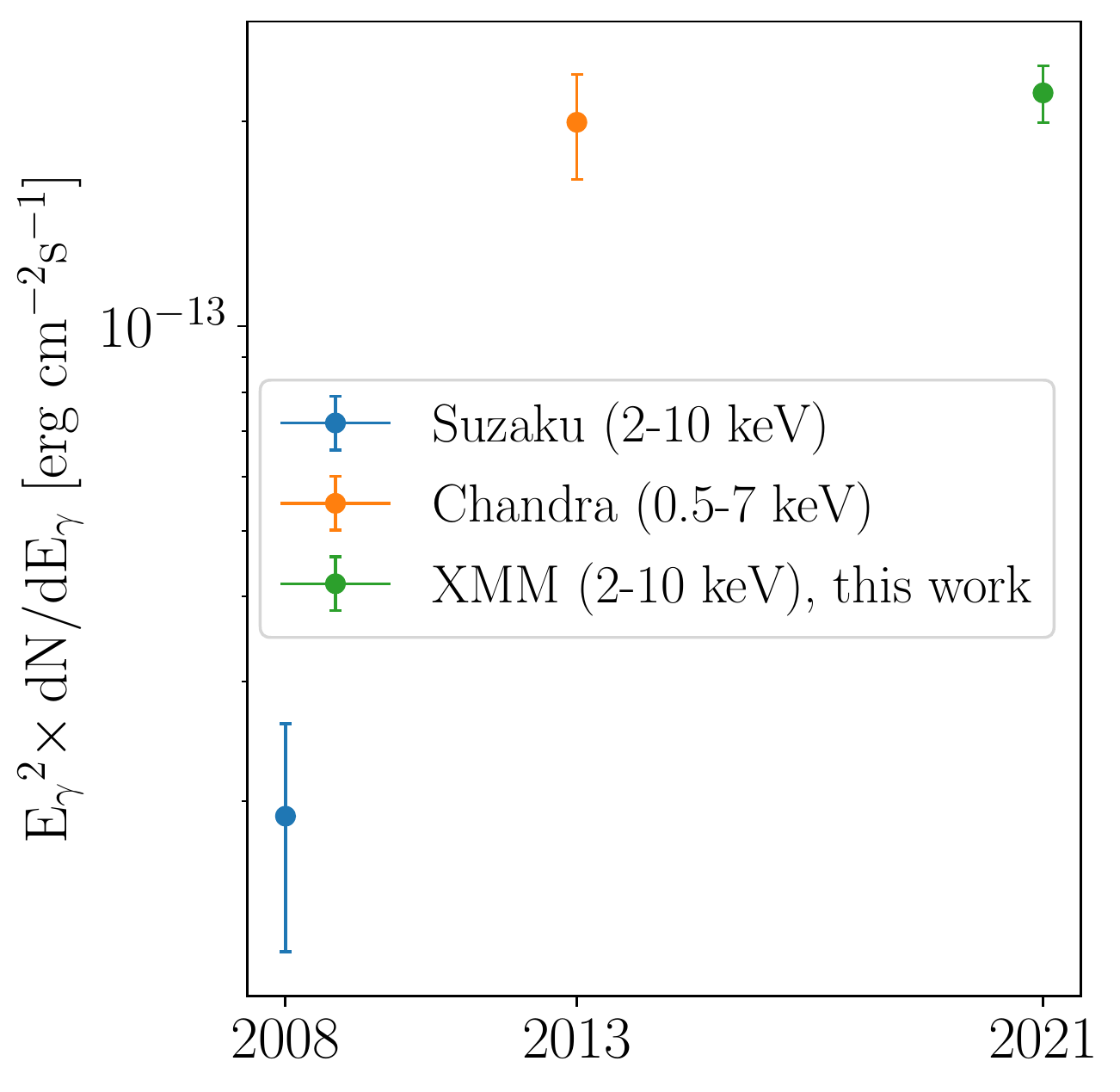}
    \caption{Integral de-absorbed X-ray flux of \Suzaku src B as measured by \Suzaku \citepalias{fujinaga}, Chandra (\mbox{2CXO J170157.2-421026})~\citep{Evans_2010} and \XMM (this work).}
    \label{variab}
\end{figure}

\begin{figure}
    \centering
    \includegraphics[width=\linewidth]{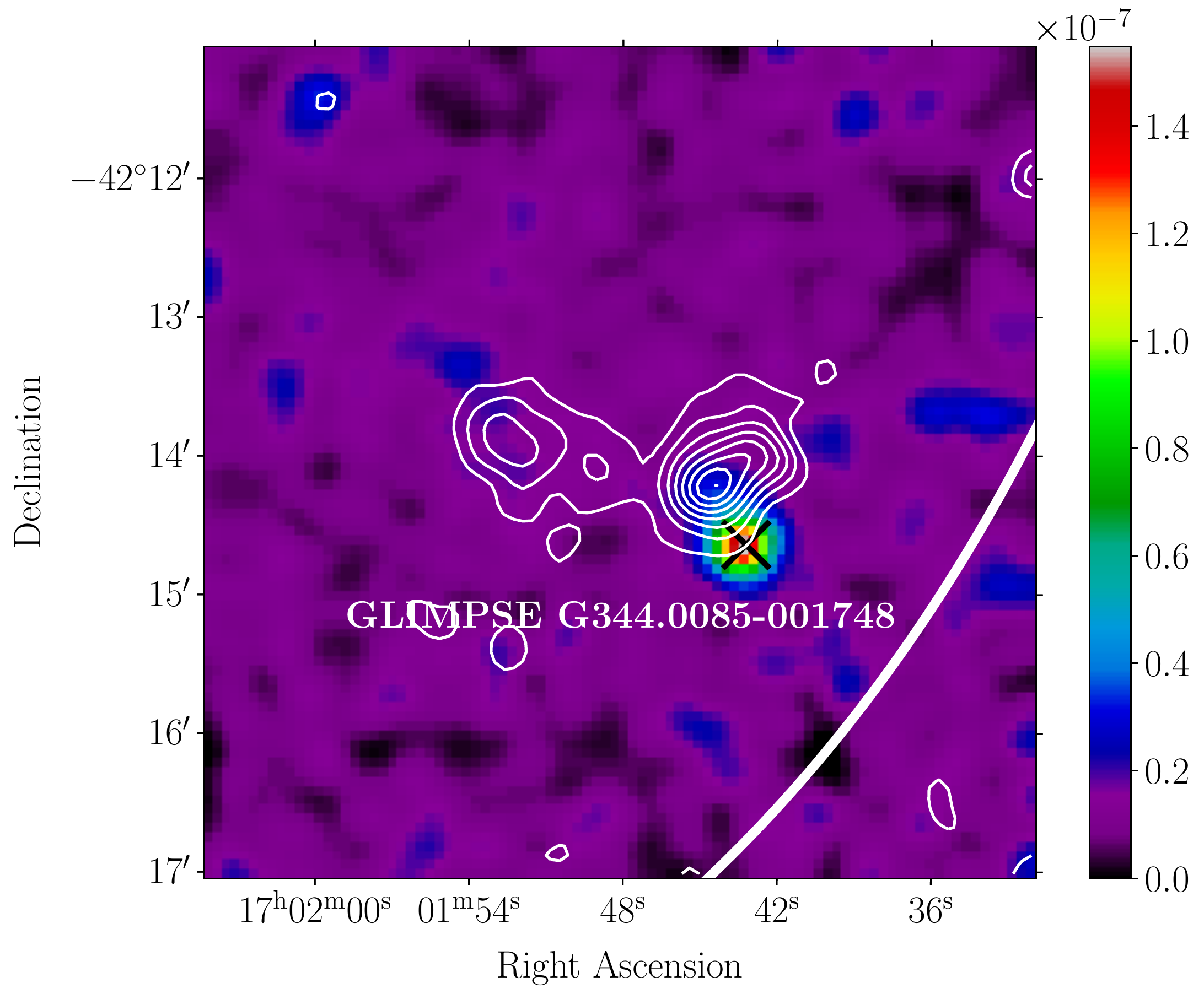}
    \caption{\XMM flux image in the $0.5-2\,$keV  energy band (color bar in units of 10$^{-7}$ ph cm$^{-2}$ s$^{-1}$ per $4''\times4''$ pixel) with white contours  showing the $2-10\,$keV data in the same region. More details are provided in the text.   }
    \label{fig:diffuse_low_energy}
\end{figure}
\begin{figure*}
    \centering
    \includegraphics[width=\linewidth]{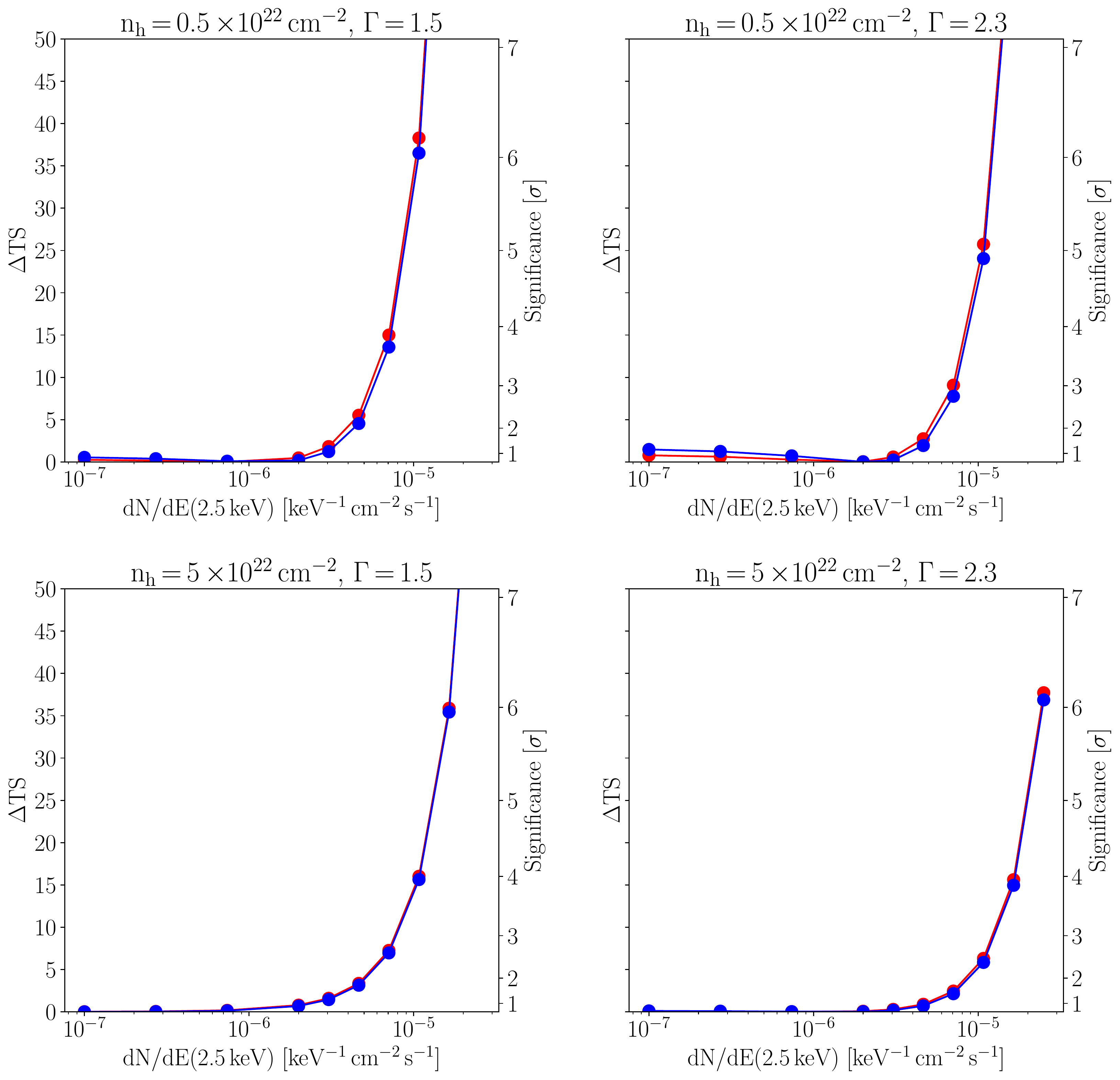}
    \caption{This figure demonstrates the successful crosscheck of the likelihood profile estimation performed with our custom Gammapy-based code (red) and the standard \emph{Sherpa} tool (blue). For each combination of assumed n$_\text{h}$ and $\Gamma$, the results obtained with the two methods are completely compatible.}
    \label{fig:validation}
\end{figure*}

\begin{table}

\centering
\begin{tabular}{ cc|ccc }
 n$_\text{h}$  & $\Gamma$ & \multicolumn{3}{c}{Flux upper limit}  \\
$[\times10^{22}\,\text{cm}^{-2}]$ &  & $1\sigma$ & $2\sigma$ & $3\sigma$ \\
\hline

 0.5 & 1.5 & 3.3 & 5.4& 7.6\\
 0.5  & 2.3 & 2.8 & 4.2& 5.7\\
 5  & 1.5 &2.4 & 5.2& 8.1\\
 5  & 2.3 &2.7 & 4.8& 6.7\\
 \hline
 
\end{tabular}
\caption{Integral de-absorbed flux (see equation~\ref{eflux})  upper limits in the \mbox{$2-10$ keV} energy band for the \jsevA region, in units of $10^{-14}\,\TEV\CMminustwo\SECminusone$, as a function of the assumed absorption column density (n$_\text{h}$) and spectral index ($\Gamma$).}
 \label{FUL}
\end{table}

\begin{table}
\centering
\begin{tabular}{ cc|ccc }

 n$_\text{h}$  & E$_\text{cut}^e$ & \multicolumn{3}{c}{Upper limit on B[$\mu G$]}  \\
$[\times10^{22}\,\text{cm}^{-2}]$ & [TeV] & $1\sigma$ & $2\sigma$ & $3\sigma$ \\
\hline

 0.5 & 100 & 0.94& 1.20& 1.48\\
 0.5  & 1000 &0.51 & 0.80& 1.05\\
 5  & 100 & 1.00& 1.45& 1.85\\
 5  & 1000 & 0.51& 0.92& 1.30\\
 \hline
 
\end{tabular}
\caption{Magnetic field upper limits in the \jsevA region, assuming a one-zone leptonic model, as a function of the assumed absorption column density (n$_\text{h}$) and intrinsic cutoff of the putative electron population(E$_\text{cut}^e$).}
 \label{BUL}
\end{table}

\end{document}